\documentclass[10pt]{article}
\usepackage[tbtags]{amsmath}
\usepackage[linktocpage]{hyperref}
\hypersetup{colorlinks, linkcolor={blue!0!black}, citecolor={blue!0!black}, urlcolor={blue!80!black}}
\usepackage{amssymb,amsthm}
\usepackage{amsfonts}
\usepackage{graphicx}
\graphicspath{{figs/}}
\usepackage[left=2.5cm,right=2.2cm,top=0.5cm,bottom=1.3cm,includeheadfoot]{geometry}
\usepackage{indentfirst}
\usepackage{color}
\usepackage{cite}
\usepackage{mathtools}
\usepackage{tikz-cd}

\usepackage[normalem]{ulem}

\newcommand{\Tr}{\mbox{Tr}}
\newcommand{\be}{\begin{equation}}
\newcommand{\ee}{\end{equation}}
\newcommand{\de}{\mbox{d}}

\numberwithin{equation}{section}

\renewcommand*{\thefootnote}{\fnsymbol{footnote}}
\begin{document}
	\begin{center}
		{\Large\bf Perturbations of bimetric gravity on most general spherically symmetric spacetimes\\
		}
		\vskip 5mm
		{\large
			David Brizuela$^{1,}$\footnote{e-mail address: {\tt david.brizuela@ehu.eus}},
			Marco de Cesare$^{2,3,}$\footnote{e-mail address: {\tt marco.decesare@na.infn.it}}, and Araceli Soler Oficial$^{1,}$\footnote{e-mail address: {\tt araceli.soler@ehu.eus}}}
		\vskip 3mm
		{\sl $^{1}$Department of Physics and EHU Quantum Center, University of the Basque Country UPV/EHU,\\
			Barrio Sarriena s/n, 48940 Leioa, Spain}\\\vskip 1mm
		{\sl $^2$ Scuola Superiore Meridionale, Largo San Marcellino 10, 80138 Napoli, Italy}\\\vskip 1mm
        {$^3$ INFN, Sezione di Napoli, Italy}
	\end{center}
	
\setcounter{footnote}{0}
\renewcommand*{\thefootnote}{\arabic{footnote}}
 
	\begin{abstract}
We present a formalism to study linear perturbations of bimetric gravity on any spherically symmetric background, including dynamical spacetimes. The setup is based on the Gerlach-Sengupta formalism for general relativity. Each of the two background metrics is written as a warped product between a two-dimensional Lorentzian metric and the round metric of the two-sphere. The different perturbations are then decomposed in terms of tensor spherical harmonics, which makes the two polarity (axial and polar) sectors decouple. In addition, a covariant notation on the Lorentzian manifold is used so that all expressions are valid for any coordinates. In this theory, there are seven physical propagating degrees of freedom, which, as compared to the two degrees of freedom of general relativity, makes the dynamics much more intricate. In particular, we discuss the amount of gauge and physical degrees of freedom for different polarities and multipoles. Finally, as an interesting application, we analyze static nonbidiagonal backgrounds and derive the corresponding perturbative equations.
	\end{abstract}

\tableofcontents

\section{Introduction}

Bimetric theory, as formulated in Refs. \cite{Hassan:2011vm, Hassan:2011tf}, is a modified gravity theory that extends general relativity (GR) by considering the existence of two coupled dynamical metrics. In particular, the corresponding interaction potential has a certain specific form in order to ensure the absence of the Boulware-Deser ghost. In this context, it has been shown that bimetric gravity is stable and well behaved in certain regions of parameter space \cite{Fasiello:2013woa}. The relevance of this theory lies in its potential to address cosmological questions, such as the accelerated expansion of the Universe and the nature of dark matter. In this sense, it is known that viable cosmological solutions that fit the expansion history of the accelerating Universe exist \cite{Volkov:2011an, Volkov:2012zb, DeFelice:2014nja, Akrami:2015qga} and that the massive spin-two field can play the role of dark matter \cite{Aoki:2014cla, Bernard:2014psa, Blanchet:2015sra, Blanchet:2015bia}. Moreover, constraints on the parameters of the theory have been derived through observational  \cite{vonStrauss:2011mq, Caravano:2021aum, Hogas:2021lns, Hogas:2021saw} and analytical \cite{Hogas:2021fmr} methods.

The stability and viability of black-hole solutions within bimetric gravity have been widely addressed in the literature \cite{Babichev:2015xha, Torsello:2017cmz}. Static and spherically symmetric black-hole solutions split into two different branches \cite{Comelli:2011wq}. In the first branch, a coordinate system exists in which the two metrics can be simultaneously diagonalized. These types of solutions are known as \textit{bidiagonal} solutions. However, this is not possible in general, and therefore there exists a second branch of \textit{nonbidiagonal} solutions. For bidiagonal solutions, the corresponding equations of motion cannot, in general, be solved analytically. Nevertheless, some exact black-hole solutions have been found in spherical symmetry \cite{Babichev:2014tfa, Babichev:2014fka}, all of them corresponding to the standard GR solutions (i.e., Schwarzschild, Schwarzschild-de Sitter, and Schwarzschild-anti-de Sitter). In fact, using analytical and numerical techniques, in Ref.~\cite{Torsello:2017cmz} it was shown that, within the bidiagonal ansatz, all black-hole solutions with flat or de Sitter asymptotics correspond to GR solutions, with both metrics being conformal (see also Ref.~\cite{Volkov:2012wp}). This, together with the fact that it is known that bidiagonal solutions where both metrics are Schwarzschild are dynamically unstable \cite{Babichev:2013una, Brito:2013wya, Babichev:2014oua, Torsello:2017cmz}, suggests that static and spherically symmetric bidiagonal solutions cannot represent the end point of gravitational collapse \cite{Torsello:2017cmz}. In contrast, with a nonbidiagonal ansatz, both metrics obey the Einstein equations and thus correspond to standard GR solutions \cite{Comelli:2011wq}. However, the correspondence with black holes in GR only holds at the background level, and it is broken by perturbations. In particular, previous work \cite{Babichev:2015zub} proved the stability of a particular subclass of nonbidiagonal static black-hole solutions against generic linear perturbations, although not for general nonbidiagonal black holes.

In this work, we present the equations for linear perturbations around a completely general (including dynamical) spherically symmetric background within bimetric theory. To this end, we use the Gerlach-Sengupta formalism \cite{PhysRevD.19.2268, PhysRevD.22.1300, Gundlach:1999bt}, based on a 2+2 decomposition of the spacetime separating the spherical symmetry orbits from a general two-dimensional Lorentzian manifold. Making use of the tensor spherical harmonics, this allows us to use a compact and covariant description of the perturbative equations both on the Lorentzian manifold and on the two-sphere, which is valid for any coordinate choice. As noted, the formalism describes the evolution of the perturbations on any spherical static black-hole or star \cite{Aoki:2016eov} backgrounds, but could also be used in the dynamical case to study, for instance, the stability during a spherically symmetric gravitational collapse \cite{Hogas:2019cpg, Kocic:2020pnm}. Here, as an interesting application, we specialize the obtained equations to a general nonbidiagonal background with a static physical metric. In this case, the analytical form of the background can be solved up to a function that satisfies a nonlinear partial differential equation \cite{Volkov:2012wp, Volkov2015}. In order to perform the computations of the present paper, we have made extensive
use of the different packages of the {\it xAct} project \cite{xAct} for Wolfram Mathematica, and particularly of {\it xPert} \cite{Brizuela:2008ra}.

The remainder of this paper is organized as follows. In Sec.~\ref{Sec:ReviewBimetric} we present the formulation of linear perturbations of bimetric gravity. In Sec.~\ref{Sec:SSBackground} we take spherically symmetric background spacetimes and introduce the 2+2 decomposition characteristic of the Gerlach-Sengupta formalism. Then, in Sec.~\ref{Sec:Harmonics}, we decompose the metric perturbations in tensor spherical harmonics. In Sec.~\ref{Sec:First-orderEOM} we discuss the gauge freedom of the theory, and obtain the equations for the linear perturbations for any two spherically symmetric background metrics, both for the axial and the polar sectors. We specialize these expressions to nonbidiagonal backgrounds in Sec.~\ref{Sec:StaticBackground}. Finally, in Sec.~\ref{Sec:Conclusion}, we review and discuss the main results of the paper.\\

\noindent\textbf{Notations and conventions:} We assume the metric signature $(-+++)$ and units with the speed of light $c=1$. The symmetrization of indices is denoted by round brackets and includes a factor of $1/2$, that is, $T_{(ab)}\coloneqq\frac{1}{2}(T_{ab}+T_{ba})$.

\section{Linear perturbations of bimetric gravity}\label{Sec:ReviewBimetric}

The bimetric gravity theory proposed by Hassan and Rosen \cite{Hassan:2011zd} is based on the existence of two dynamical and nonlinearly interacting metrics, $\tilde g_{\mu\nu}$ and $\tilde f_{\mu\nu}$, on the four-dimensional spacetime manifold. The action is given by the linear combination of the Einstein-Hilbert term for each metric, complemented with a coupling term
\be\label{Eq:BimetricAction}
S_{\rm\scriptscriptstyle HR}=\frac{M_g^2}{2} \int \de^4 x \sqrt{-\tilde g}\, {\cal R}^{(\tilde g)}
+\frac{M_f^2}{2} \int \de^4 x \sqrt{-\tilde f}\, {\cal R}^{(\tilde f)}-m^2 M_g^2 \int \de^4x \sqrt{-\tilde g}\,\sum_{n=0}^4 \beta_n e_n(\tilde{\mathbb{S}})~,
\ee
where ${\cal R}^{(\tilde g)}$ and ${\cal R}^{(\tilde f)}$ are the Ricci scalars of the metrics $\tilde g_{\mu\nu}$ and $\tilde f_{\mu\nu}$, respectively.  The coupling constants $M_g$, $M_f$, and $m$ have dimensions of mass, while the $\beta_n$ are dimensionless. Finally, the $e_n$ are symmetric polynomials of scalar combinations of the matrix \cite{Hassan:2017ugh}
\be\label{Eq:S-matrix}
\tilde{\mathbb{S}}^\mu{}_{\nu}=\sqrt{\tilde g^{\mu\alpha} \tilde f_{\alpha\nu}}~,
\ee
and are explicitly defined as \cite{Hassan:2011hr,Bernard:2015mkk}
\begin{subequations}\label{Eq:SymmetricPoly_Defs}
\begin{align}
e_0 (\tilde{\mathbb{S}})&=1~,\\
e_1 (\tilde{\mathbb{S}})&=\Tr[\tilde{\mathbb{S}}] ~,\\
e_2 (\tilde{\mathbb{S}})&=\frac{1}{2}\left(\Tr[\tilde{\mathbb{S}}]^2 -  \Tr[\tilde{\mathbb{S}}^2]\right)~,\\
e_3 (\tilde{\mathbb{S}})&=\frac{1}{6}\left(\Tr[\tilde{\mathbb{S}}]^3 - 3\Tr[\tilde{\mathbb{S}}] \Tr[\tilde{\mathbb{S}}^2]+2\Tr[\tilde{\mathbb{S}}^3]\right)~,\\
e_4 (\tilde{\mathbb{S}})&=\frac{1}{24}\left(\Tr[\tilde{\mathbb{S}}]^4 - 6\Tr[\tilde{\mathbb{S}}]^2 \Tr[\tilde{\mathbb{S}}^2]+3\Tr[\tilde{\mathbb{S}}^2]^2 + 8 \Tr[\tilde{\mathbb{S}}] \Tr[\tilde{\mathbb{S}}]^3 - 6\Tr[\tilde{\mathbb{S}}^4]  \right)~,
\end{align}
\end{subequations}
with $\Tr[\tilde{\mathbb{S}}]=\tilde{\mathbb{S}}^\mu{}_{\mu}$. For a $d\times d$ matrix $\tilde{\mathbb{S}}$, $e_n(\tilde{\mathbb{S}})=0$ for any $n>d$ and $e_d(\tilde{\mathbb{S}})=\det{(\tilde{\mathbb{S}})}$. Therefore, one could also write $e_4(\tilde{\mathbb{S}})=\det{(\tilde{\mathbb{S}})}$. By the relation
\begin{equation}\label{Eq:Relationg-f-action}
    \sqrt{-\tilde{g}}\,e_n(\mathbb{\tilde S})=\sqrt{-\tilde{f}}\,e_{4-n}(\mathbb{\tilde S}^{-1})~,
\end{equation}
it is straightforward to see that the bimetric action \eqref{Eq:BimetricAction} is invariant under the simultaneous replacements
\begin{equation}\label{Eq:Exchangeg-f}
    \tilde{g}\leftrightarrow \tilde{f},\quad \beta_n\leftrightarrow \beta_{4-n},\quad M_g \leftrightarrow M_f, \quad m^2\leftrightarrow m^2M_g^2/M_f^2,
\end{equation}
which means that both metrics are treated on the same footing in the pure gravity theory. However, such a symmetry is broken by matter fields, which typically are only coupled to a single metric \cite{deRham:2014naa, Yamashita:2014fga} (see also the review \cite{Schmidt-May:2015vnx}). Therefore, here we will also assume that matter sources couple only to the  metric $g_{\mu\nu}$, and are described by the corresponding stress-energy tensor $\mathcal{T}_{\mu\nu}$. This leads to the equations of motion
\begin{subequations}\label{Eq:BigravityEom_matter}
\begin{align}\label{Eq:BigravityEom_matter-g}
G_{\mu\nu}^{(\tilde g)}+m^2V_{\mu\nu}^{(\tilde g)}&=\frac{1}{M_g^2}\mathcal{T}_{\mu\nu}~,\\\label{Eq:BigravityEom_matter-f}
G_{\mu\nu}^{(\tilde{f})}+\frac{m^2}{\alpha^2}V_{\mu\nu}^{(\tilde f)} &=0~,
\end{align}
\end{subequations}
where $G_{\mu\nu}^{(\tilde g)}$ and $G_{\mu\nu}^{(\tilde f)}$ are the Einstein tensors of the corresponding metrics, while $\alpha\coloneqq M_f/M_g$ measures the ratio between the gravitational couplings. The interaction between the two metrics is encoded in the potential
\begin{subequations}\label{Eq:InteranctionPotentials}
\begin{align}
    V_{\mu\nu}^{(\tilde g)}&:=\frac{1}{2}\sum_{i=0}^3(-1)^i\beta_i\left[\tilde g_{\mu\rho}(Y_{(i)})^{\rho}_{\,\,\nu}(\tilde{\mathbb{S}})+\tilde g_{\nu\rho}(Y_{(i)})^{\rho}_{\,\,\mu}(\tilde{\mathbb{S}})\right]~,\\
    V_{\mu\nu}^{(\tilde f)}&:=\frac{1}{2}\sum_{i=0}^3(-1)^i\beta_{4-i}\left[\tilde f_{\mu\rho}(Y_{(i)})^{\rho}_{\,\,\nu}(\tilde{\mathbb{S}}^{-1})+\tilde f_{\nu\rho}(Y_{(i)})^{\rho}_{\,\,\mu}(\tilde{\mathbb{S}}^{-1})\right]~,
\end{align}
\end{subequations}
where the matrices $Y_{(i)}$ read as
 \begin{equation}\label{Eq:YMatrices}
    (Y_{(i)})^{\rho}_{\,\,\nu}(\tilde{\mathbb{S}}):= \sum_{k=0}^i(-1)^k(\tilde{\mathbb{S}}^{i-k})^{\rho}_{\,\,\nu}e_k(\tilde{\mathbb{S}})~.
\end{equation}

In the following we will consider the effective stress-energy tensors
\begin{align}
   t_{\mu\nu}^{(\tilde g)}&\coloneqq\frac{1}{8\pi M_g^2}\mathcal{T}_{\mu\nu}-\frac{m^2}{8\pi}V_{\mu\nu}^{(\tilde g)}~,\label{Eq:SE-Tensor-g}\\
   t_{\mu\nu}^{(\tilde f)}&\coloneqq-\frac{m^2}{8\pi\alpha^2}V_{\mu\nu}^{(\tilde f)}~,
\end{align}
so that the equations of motion \eqref{Eq:BigravityEom_matter} formally take the same form as the Einstein equations,
\begin{subequations}\label{Eq:BigravityEom}
\begin{align}
G_{\mu\nu}^{(\tilde g)}&=8\pi t^{(\tilde g)}_{\mu\nu}~,\\
G_{\mu\nu}^{(\tilde f)}&=8\pi t^{(\tilde f)}_{\mu\nu}~.
\end{align}
\end{subequations}

Now, in order to perform a perturbative analysis of the theory, we write
\begin{subequations}\label{Eq:DefMetricPerturbations}
\begin{align}
 \tilde g_{\mu\nu}= g_{\mu\nu}+ h^{(g)}_{\mu\nu}~,\\
 \tilde f_{\mu\nu}= f_{\mu\nu}+ h^{(f)}_{\mu\nu}~,
\end{align}
\end{subequations}
where the metrics $g_{\mu\nu}$ and $f_{\mu\nu}$ are exact solutions of the equations \eqref{Eq:BigravityEom} and will be referred to as the {\it background}. In turn, $h^{(g)}_{\mu\nu}$ and $h^{(f)}_{\mu\nu}$ encode the {\it perturbations} and will be assumed to be small. That is,  in order to obtain their equations of motion, one simply substitutes the ansatz \eqref{Eq:DefMetricPerturbations} into \eqref{Eq:BigravityEom}, and regards any term quadratic in the perturbations as negligible.

Let us define the operator $\Delta$ as providing the linear term in $h^{(g)}_{\mu\nu}$ and $h^{(f)}_{\mu\nu}$ of any object; for instance, $t^{(\tilde g)}_{\mu\nu}=t^{(g)}_{\mu\nu}+\Delta[t^{(g)}_{\mu\nu}]$. In this way, the linear equations of motion for $h^{(g)}_{\mu\nu}$ and $h^{(f)}_{\mu\nu}$ can be written as
\begin{subequations}\label{Eq:BigravityEom_Perturbed}
\begin{align}
\Delta[G_{\mu\nu}^{(g)}]&=8\pi \Delta[t^{(g)}_{\mu\nu}]~,\\
\Delta[G_{\mu\nu}^{(f)}]&=8\pi \Delta[t^{(f)}_{\mu\nu}]~.
\end{align}
\end{subequations}
The left-hand side are the perturbations of the Einstein tensor of each metric,  whose form is well known,
\begin{equation}
2 \Delta[G^{(g)}_{\mu\nu}]=h^{(g)\alpha}_{\,\,\mu}{}_{;\nu\alpha}+h^{(g)\alpha}_{\,\,\nu}{}_{;\mu\alpha}-h^{(g)}_{\mu\nu}{}^{;\alpha}{}_{\alpha}-h^{(g)\alpha}_{\,\,\alpha}{}_{;\mu\nu}-h^{(g)}_{\mu\nu}\mathcal{R}^{(g)}-g_{\mu\nu}\left(h^{(g)\alpha\beta}{}_{;\alpha\beta}-h^{(g)\alpha}_{\,\,\alpha}{}^{;\beta}{}_{\beta}-h^{(g)\alpha\beta}\mathcal{R}^{(g)}_{\alpha\beta}\right)~,
\end{equation}
where the semicolon ``~$;$~'' denotes the covariant derivative associated to $g_{\mu\nu}$, and $\mathcal{R}^{(g)}_{\alpha\beta}$ is its Ricci tensor. The perturbation of the Einstein tensor of the metric $\tilde{f}_{\mu\nu}$ can be computed analogously. Therefore, the nontrivial part of the present computation will be to obtain the linear version of the effective stress-energy tensors $t^{(\tilde g)}_{\mu\nu}$ and $t^{(\tilde f)}_{\mu\nu}$. In particular, this requires us to compute the perturbation of the matrix $\tilde{\mathbb{S}}^\mu{}_\nu$. By definition, we have
\begin{equation}
 \tilde{\mathbb{S}}^\mu{}_\alpha \tilde{\mathbb{S}}^\alpha{}_\nu=\tilde g^{\mu\alpha} \tilde f_{\alpha\nu}~.
\end{equation} 
Replacing the expansions \eqref{Eq:DefMetricPerturbations} and $\tilde{\mathbb{S}}^\mu{}_\alpha=\mathbb{S}^\mu{}_\alpha+\Delta[\mathbb{S}^\mu{}_\alpha]$ in this expression, the term linear in perturbations yields
\begin{equation}
    \qquad   \mathbb{S}^{\mu}_{\,\,\alpha}\Delta[\mathbb{S}^{\alpha}_{\,\,\nu}]
    +\Delta[ \mathbb{S}^{\mu}_{\,\,\alpha}] \mathbb{S}^{\alpha}_{\,\,\nu}=g^{\mu\alpha}h^{(f)}_{\alpha\nu}-g^{\mu\alpha}h^{(g)}_{\alpha\beta}g^{\beta\sigma}f_{\sigma\nu}~,
    \end{equation}
which can be rewritten as
    \begin{equation}\label{Eq:PertS}
    (\mathbb{S}^\mu_{\,\,\alpha }\delta^\alpha_{\,\,\rho}\delta^\sigma_{\,\,\nu}+\mathbb{S}^\alpha_{\,\,\nu}\delta^\mu_{\,\,\rho}\delta^\sigma_{\,\,\alpha})\Delta[\mathbb{S}^\rho_{\,\,\sigma}]=g^{\mu\alpha}h^{(f)}_{\alpha\nu}-g^{\mu\alpha}h^{(g)}_{\alpha\beta}g^{\beta\sigma}f_{\sigma\nu}~.
    \end{equation}
Hence, in order to obtain $\Delta[\mathbb{S}^\rho_{\,\,\sigma}]$ explicitly in terms of $h^{(g)}_{\mu\nu}$ and $h^{(f)}_{\mu\nu}$, one would need to compute the inverse of the expression in brackets above.  Although this does not seem feasible for generic backgrounds, on a spherically symmetric background the problem can be simplified by decomposing $\Delta[\mathbb{S}^\rho_{\,\,\sigma}]$ in a basis of tensor spherical harmonics, as we will show in Sec.~\ref{Sec:decomposeDeltaS}.

Before we move on to analyze perturbations around specific backgrounds, let us comment on the gauge freedom and the number of propagating degrees of freedom. In vacuum GR the only dynamical field is the metric, which, being a rank-two symmetric tensor field, in principle encodes ten independent local degrees of freedom. However, there are eight first-class constraints: four corresponding to the generators of diffeomorphisms (the so-called Hamiltonian and diffeomorphism constraints), and four more corresponding to the vanishing of the conjugate momenta of lapse and shift. Each of these constraints removes one degree of freedom (i.e., two phase-space dimensions per spacetime point), which leaves a total of two propagating degrees of freedom, corresponding to two independent polarizations for the graviton. Considering two decoupled copies of GR, hence with two independent groups of diffeomorphisms (each acting independently on a single metric sector), both the number of degrees of freedom and constraints would double, and one would have four propagating degrees of freedom. However, when the two metrics are coupled, a set of four first-class constraints of the system is removed, due to the now common diffeomorphism invariance. For generic choices of the potential describing the interaction between the two metrics, this would lead to eight propagating degrees of freedom in total. In the Hassan-Rosen theory though, given by the action \eqref{Eq:BimetricAction}, the coupling term is chosen in such a way that there appears a couple of second-class constraints \cite{Hassan:2011hr, Hassan:2011tf, Hassan:2011ea} that remove one degree of freedom, the so-called Boulware-Deser ghost, leaving seven propagating degrees of freedom \cite{Alexandrov:2013rxa, deRham:2014zqa, Hassan:2018mbl}.

\section{Spherically symmetric background}\label{Sec:SSBackground}

Any four-dimensional spherically symmetric manifold is given as a direct product ${\cal M}^2\times {\cal S}^2$, where ${\cal M}^2$ is a two-dimensional Lorentzian manifold and ${\cal S}^2$ is the two-sphere. The background metric tensors can then be written in block-diagonal form,
\begin{align}
    g_{\mu\nu}(x^\lambda)\de x^\mu\de x^\nu & =g_{AB}(x^D)\de x^A\de x^B+r_g^2(x^D)\gamma_{ab}(x^d)\de x^a\de x^b~,\\
    f_{\mu\nu}(x^\lambda)\de x^\mu\de x^\nu & = f_{AB}(x^D)\de x^A\de x^B+r_f^2(x^D)\gamma_{ab}(x^d)\de x^a\de x^b~,
\end{align}
where Greek indices take values from 0 to 3, capital Latin indices from 0 to 1, and lowercase Latin indices run from 2 to 3. The tensor
\begin{equation}
    \gamma_{ab}(x^d)\de x^a\de x^b=\de \theta^2+\sin^2\theta\de \varphi^2
\end{equation}
is the unit metric on the two-sphere, while $g_{AB}$ and $f_{AB}$ are Lorentzian metrics in ${\cal M}^2$. With this decomposition, the matrix $\mathbb{S}^{\mu}_{\,\,\,\nu}$ defined by Eq.~\eqref{Eq:S-matrix} is also diagonal by blocks with $\mathbb{S}^a_{\,\,\,b}= \frac{r_f}{r_g}\,\delta^a_{\,\,\,b}$. For future convenience, we define the determinant of the block in the $\mathcal{M}^2$ sector as $\mathcal{D}\coloneqq\det{(\mathbb{S}^A{}_B)}$ and the ratio between the two area radii as $\omega\coloneqq r_f/r_g$.

The nonvanishing components of the bimetric equations \eqref{Eq:BigravityEom} for any general spherically symmetric spacetimes, which define our gravitational background, read as
    \begin{align}\label{Eq:Background-eom-1}
        G_{AB}^{(i)} &=\left(\frac{1}{r_i^2}+3v^{(i)D}{v^{(i)}}_{\hspace{-0.07cm}D}+2\overset{(i)}{\nabla}_{\hspace{-0.07cm}D}  v^{(i)D}\right)\mathfrak{g}^{(i)}_{AB}-2\left({v^{(i)}}_{\hspace{-0.07cm}A}{v^{(i)}}_{\hspace{-0.07cm}B}+\overset{(i)}{\nabla}_{\hspace{-0.07cm}B}{v^{(i)}}_{\hspace{-0.07cm}A}\right)=8\pi t^{(i)}_{AB}~,\\
            \frac{\gamma^{ab}G_{ab}^{(i)}}{r_i^2} &=-R^{(i)}+2v^{(i)A}{v^{(i)}}_{\hspace{-0.07cm}A}+2\overset{(i)}{\nabla}_{\hspace{-0.07cm}A}v^{(i)A}=8\pi Q_i\label{Eq:Background-eom}~,
    \end{align}
where we have introduced the label $i\in \{g,~f\}$ to write collectively the two metric sectors, with $\mathfrak{g}^{(g)}_{AB}=g_{AB}$ and $\mathfrak{g}^{(f)}_{AB}=f_{AB}$. In addition, we have defined the vector fields
\begin{equation}
v_A^{(i)}=\frac{\partial_Ar_i}{r_i}~,
\end{equation}
and $Q_i\coloneqq\frac{\gamma^{ab}}{r_i^2}t^{(i)}_{ab}$ on ${\cal M}^2$, while $R^{(i)}$ stands for the Ricci scalar of the corresponding two-dimensional metrics $g_{AB}$ and $f_{AB}$. We have also introduced $\overset{(g)}{\nabla}$ and $\overset{(f)}{\nabla}$ as the covariant derivatives of the metrics $g_{AB}$ and $f_{AB}$, respectively. This notation will be used throughout the paper. Moreover, in expressions with a label $i$, and wherever repeated capital Latin indices appear, the ensuing contraction should be understood as being performed with the corresponding metric $\mathfrak{g}^{(i)}_{AB}$.

Finally, the components of the effective stress-energy tensors on $\mathcal{M}^2$ are explicitly given by
\begin{align}
\begin{split}
    t_{AB}^{(g)} = & -\frac{m^2}{8\pi}\Big[\left(\beta_0+2\omega\beta_1+\omega^2\beta_2+(\beta_1+2\omega\beta_2+\omega^2\beta_3)\mathbb{S}^B_{\,\,\,B}+(\beta_2+2\omega\beta_3)\mathcal{D}\right)\,g_{AB}\\
    & -\big(\left(\beta_1+2\omega\beta_2+\omega^2\beta_3+(\beta_2+2\omega\beta_3)\mathbb{S}^E_{\,\,\,E}+\beta_3\mathcal{D}\right)\delta^C_{\,\,\,D}-(\beta_2+2\omega\beta_3+\beta_3\mathbb{S}^F_{\,\,\,F})\mathbb{S}^C_{\,\,\,D}\\
    & +\beta_3\mathbb{S}^C_{\,\,\,E}\mathbb{S}^E_{\,\,\,D}\big)g_{C(A}\mathbb{S}^D_{\,\,\,B)}\Big]+\frac{\mathcal{T}_{AB}}{8\pi M_g^2}~,
\end{split}\\
\begin{split}
    t_{AB}^{(f)} = & -\frac{m^2}{8\pi\alpha^2\omega^2}\Big[\left(\omega^2\beta_4+2\omega\beta_3+\beta_2+(\omega^2\beta_3+2\omega\beta_2+\beta_1)(\mathbb{S}^{-1})^B_{\,\,\,B}+(\omega^2\beta_2+2\omega\beta_1)\mathcal{D}^{-1}\right)\,f_{AB}\\
    & -\big(\left(\omega^2\beta_3+2\omega\beta_2+\beta_1+(\omega^2\beta_2+2\omega\beta_1)(\mathbb{S}^{-1})^E_{\,\,\,E}+\omega^2\beta_1\mathcal{D}^{-1}\right)\delta^C_{\,\,\,D}\\
    & -(\omega^2\beta_2+2\omega\beta_1+\omega^2\beta_1(\mathbb{S}^{-1})^F_{\,\,\,F})(\mathbb{S}^{-1})^C_{\,\,\,D}+\omega^2\beta_1(\mathbb{S}^{-1})^C_{\,\,\,E}(\mathbb{S}^{-1})^E_{\,\,\,D}\big)f_{C(A}(\mathbb{S}^{-1})^D_{\,\,\,B)}\Big]~,
\end{split}
\end{align}
while the traces of the angular components read as
\begin{align}
    Q_g&= -\frac{m^2}{4\pi}\left[\left(\beta_0+\omega \beta_1\right) +\left(\beta_1+\omega \beta_2\right) \mathbb{S}^A_{\,\,\,A}+\left(\beta_2+\omega \beta_3\right) \mathcal{D}\right]+Q_{m}~,\\
    Q_f &= -\frac{m^2}{4\pi \alpha^2\omega^2}\left[\left(\omega^2\beta_4+\omega \beta_3\right) +\left(\omega^2\beta_3+\omega \beta_2\right) (\mathbb{S}^{-1})^A_{\,\,\,A}+\left(\omega^2\beta_2+\omega \beta_1\right) \mathcal{D}^{-1}\right]~,
\end{align}
where we have defined the contribution from the matter sector as $Q_m\coloneqq\frac{\gamma^{ab}\mathcal{T}_{ab}}{8\pi M_g^2r_g^2}$.

\section{Harmonic decomposition}\label{Sec:Harmonics}

\subsection{Tensor spherical harmonics}

The usual scalar spherical harmonics $Y_l^m=Y_l^m(x^a)$ are defined as the eigenfunctions of the Laplacian operator acting on scalars,
\begin{equation}
 \gamma^{ab}Y_l^m{}_{:ab}=-l(l+1)Y_l^m~,
\end{equation}
where ``~$:$~'' is the covariant derivative associated with $\gamma_{ab}$, while $l$ and $m$
are integers such that $l\geq |m|$. These special functions form a basis on the sphere, and thus any scalar function $F=F(x^a)$ can be written as a linear combination
\begin{equation}
F(x^a)=\sum_{l=0}^\infty \sum_{m=-l}^l F_l^m Y_l^m(x^a)~,
\end{equation}
with certain complex constants $F_l^m$.

Making use of the metric $\gamma_{ab}$, its covariant derivative, and the antisymmetric tensor\footnote{The antisymmetric tensor $\epsilon_{ab}$ is defined as $\epsilon_{ab}:=\sqrt{\gamma}\,\eta_{ab}$, where $\gamma$ is the determinant of $\gamma_{ab}$ and $\eta_{ab}$ the antisymmetric symbol with $\eta_{23}=-\eta_{32}=1$.}
$\epsilon_{ab}$ on ${\cal S}^2$, it is possible to generalize this basis to tensors of any rank (see Ref. \cite{Brizuela:2006ne} for more details). For instance, a basis for vectors on the sphere is given by the two vectors $Z_l^m{}_a\coloneqq\partial_a Y_l^m$ and $X_l^m{}_a\coloneqq\epsilon_a{}^{b}Z_l^m{}_b$, which are irrotational and divergence-free, respectively. Thus, any vector $F_a(x^b)$ can be decomposed as,
\begin{equation}
F_a(x^b)=\sum_{l=1}^\infty \sum_{m=-l}^l \tilde{F}_l^m Z_l^m{}_a(x^b)+ \hat{F}_l^m X_l^m{}_a(x^b)~,
\end{equation}
where $\tilde{F}_l^m$ and $\hat{F}_l^m$ are constants. In the theory under consideration there are also rank-two tensors, for which we will use the basis $\{Z_l^m{}_{ab},\, X_l^m{}_{ab},\, \gamma_{ab} Y_l^m ,\, \epsilon_{ab} Y_l^m \}$, with $Z_l^m{}_{ab}\coloneqq Y_l^m{}_{:ab}+\frac{l(l+1)}{2}\gamma_{ab}Y_l^m$ and $X_l^m{}_{ab}\coloneqq\frac{1}{2}(X_l^m{}_{a:b}+X_l^m{}_{b:a})$ being symmetric and trace-free.

The tensor harmonics have different polarity properties and they are divided into polar (or even) and axial (or odd) polarities. In the case of scalar functions, only polar components appear, and, in order to have a more uniform notation, we will denote $Z_l^m\coloneqq Y_l^m$. In this way, all the terms multiplying a $Z$ are polar, while those multiplying an $X$ are axial. It is a well-known result in GR that different polarities decouple at the linear level, so long as the background is spherically symmetric. This holds true also in bimetric gravity, as explicitly shown below.

Finally, we note that different harmonics are defined for a different range of values of $l$. More precisely, while scalar harmonics are defined for $l\geq 0$, vector harmonics $Z_l^m{}_{a}$ and $X_l^m{}_{a}$ are exactly vanishing for $l=0$, and thus only contribute for $l\geq 1$, while tensor harmonics $Z_l^m{}_{ab}$ and $X_l^m{}_{ab}$ are nonvanishing only for $l\geq 2$.

\subsection{Decomposition of the perturbations into tensor spherical harmonics}\label{SubSec:Decomposition}

The components of the metric perturbations $h_{\mu\nu}^{(i)}$ have different tensorial rank in ${\cal S}^2$, which can be easily identified in terms of their indices. Namely, $h_{AB}^{(i)}$ is a scalar, $h_{Ab}^{(i)}$ is a vector, and $h_{ab}^{(i)}$ is a symmetric rank-two tensor. Therefore, one needs to use a suitable basis, given by tensor spherical harmonics of the appropriate rank, as explained above. In this way, we introduce the following decompositions:
\begin{align}
    h_{AB}^{(i)}(x^D,x^d) & \coloneqq\sum_{l=0}^\infty\sum _{m=-l}^l H^{(i)m}_{\hspace{0.35cm}l\,\,AB}Z^m_l~,\\
     h_{Ab}^{(i)}(x^D,x^d) & \coloneqq\sum_{l=1}^\infty\sum _{m=-l}^l\left[H^{(i)m}_{\hspace{0.35cm}l\,\,A}Z^{m}_{l\,\,\,\,b}+h^{(i)m}_{\hspace{0.35cm}l\,\,A}X^m_{l\,\,\,\,b}\right]~,\\
     h_{ab}^{(i)}(x^D,x^d) & \coloneqq\sum_{l=0}^\infty\sum _{m=-l}^lK^{(i)m}_{\hspace{0.35cm}l}r_i^2\gamma_{ab}Z^m_l
     +\sum_{l=2}^\infty\sum _{m=-l}^l\left[G^{(i)m}_{\hspace{0.35cm}l}r_i^2Z^m_{l\,\,\,\,ab}+h^{(i)m}_{\hspace{0.35cm}l}X^m_{l\,\,\,\,ab}\right]~,
\end{align}
for $i\in\{f,~g\}$. For each set of labels $({}^{(i)},l,m)$ with $l\geq 2$ there are ten new independent functions: these are encoded in a symmetric two-tensor $H^{(i)m}_{\hspace{0.35cm}l\,\,AB}$, two vectors $\{H^{(i)m}_{\hspace{0.35cm}l\,\,A},\,h^{(i)m}_{\hspace{0.35cm}l\,\,A}\}$, and three scalars $\{K^{(i)m}_{\hspace{0.35cm}l},\,G^{(i)m}_{\hspace{0.35cm}l},\,h^{(i)m}_{\hspace{0.35cm}l}\}$, all of which only depend on coordinates of $\mathcal{M}^2$. From these, seven are polar $\{H^{(i)m}_{\hspace{0.35cm}l\,\,AB},\,H^{(i)m}_{\hspace{0.35cm}l\,\,A},\,K^{(i)m}_{\hspace{0.35cm}l},\,G^{(i)m}_{\hspace{0.35cm}l}\}$ and three are axial $\{h^{(i)m}_{\hspace{0.35cm}l\,\,A},\,h^{(i)m}_{\hspace{0.35cm}l}\}$. Note that for $l=0$ we only have the four polar components $H^{(i)0}_{\hspace{0.35cm}0\,\,AB}$ and $K^{(i)0}_{\hspace{0.35cm}0}$, whereas, for $l=1$, in addition to $H^{(i)m}_{\hspace{0.35cm}1\,\,AB}$ and $K^{(i)m}_{\hspace{0.35cm}1}$, two polar components $H^{(i)m}_{\hspace{0.35cm}1\,\,A}$ and two axial components $h^{(i)m}_{\hspace{0.35cm}1\,\,A}$ are also present.

Similarly, linear perturbations of the effective stress-energy tensors read as
\begin{align}
    \Delta [t_{AB}^{(i)}](x^D,x^d) & \coloneqq\sum_{l=0}^\infty\sum _{m=-l}^l T^{(i)m}_{\hspace{0.35cm}l\,AB}Z^m_l~,\label{Eq:harmonics-tAB}\\
     \Delta [t_{Ab}^{(i)}](x^D,x^d) & \coloneqq\sum_{l=1}^\infty\sum _{m=-l}^l\left[T^{(i)m}_{\hspace{0.35cm}l\,\,\,A}Z^m_{l\,\,\,\,b}+t^{(i)m}_{\hspace{0.35cm}l\,\,\,A}X^m_{l\,\,\,\,b}\right]~,\label{Eq:harmonics-tAb}\\
     \Delta [t_{ab}^{(i)}](x^D,x^d) & \coloneqq\sum_{l=0}^\infty\sum _{m=-l}^l\widetilde{T}^{(i)m}_{\hspace{0.35cm}l}r_i^2\gamma_{ab}Z^m_l+\sum_{l=2}^\infty\sum _{m=-l}^l\left[T^{(i)m}_{\hspace{0.35cm}l}Z^m_{l\,\,\,\,ab}+t^{(i)m}_{\hspace{0.35cm}l}X^m_{l\,\,\,\,ab}\right]~.\label{Eq:harmonics-tab}
\end{align}
Note that, in the $g$-sector, according to the definition of the effective stress-energy tensors \eqref{Eq:SE-Tensor-g}, besides the perturbations of the bigravity interaction term $V^{(\tilde{g})}_{\mu\nu}$, the harmonic components defined above also include the contribution of the perturbations of the matter stress-energy tensor $\mathcal{T}_{\mu\nu}$. Hence, for future convenience, we also introduce the following decompositions
    \begin{align}
     \Delta [\mathcal{T}_{AB}](x^D,x^d) & \coloneqq\sum_{l=0}^\infty\sum _{m=-l}^l \Psi^{m}_{l\,AB}Z^m_l~,\label{Eq:harmonics-tABmatter}\\
      \Delta [\mathcal{T}_{Ab}](x^D,x^d) & \coloneqq\sum_{l=1}^\infty\sum _{m=-l}^l\left[\Psi^{m}_{l\,A}Z^m_{l\,\,\,\,b}+\psi^{m}_{l\,A}X^m_{l\,\,\,\,b}\right]~,\label{Eq:harmonics-tAbmatter}\\
     \Delta [\mathcal{T}_{ab}](x^D,x^d) & \coloneqq\sum_{l=0}^\infty\sum _{m=-l}^l\widetilde{\Psi}^{m}_{l}r_g^2\gamma_{ab}Z^m_l+\sum_{l=2}^\infty\sum _{m=-l}^l\left[\Psi^{m}_{l}Z^m_{l\,\,\,\,ab}+\psi^{m}_{l}X^m_{l\,\,\,\,ab}\right]~.\label{Eq:harmonics-tabmatter}
\end{align}

Finally, we will also need the decomposition into spherical harmonics of the components of the perturbation $\Delta[\mathbb{S}^{\mu}_{\,\,\,\nu}]$ of the matrix $\mathbb{S}$,
\begin{align}
    \Delta [\mathbb{S}^{A}_{\,\,\,B}](x^D,x^d) & \coloneqq\sum_{l=0}^\infty\sum _{m=-l}^l S^{m A}_{l\,\,\,\,\,\,B}Z^m_l~,\\
     \Delta [\mathbb{S}^{A}_{\,\,\,b}](x^D,x^d) & \coloneqq\sum_{l=1}^\infty\sum _{m=-l}^l\left[S^{mA}_{l}Z^m_{l\,\,\,\,b}+s^{mA}_{l}X^m_{l\,\,\,\,b}\right]~,\\
     \Delta [\mathbb{S}^{a}_{\,\,\,B}](x^D,x^d) & \coloneqq\frac{1}{r_g^2}\gamma^{ac}\sum_{l=1}^\infty\sum _{m=-l}^l\left[\tilde{S}^{m}_{l\,\,\,B}Z^{m}_{l\,\,\,\,c}+\tilde{s}^{\,m}_{\,l\,\,\,B}X^{m}_{l\,\,\,\,c}\right]~,\\
     \Delta [\mathbb{S}^{a}_{\,\,\,b}](x^D,x^d) & \coloneqq\gamma^{ac}\sum_{l=0}^\infty\sum _{m=-l}^l \tilde{S}^{m}_{l}\gamma_{cb}Z^m_l+\gamma^{ac}\sum_{l=2}^\infty\sum _{m=-l}^l\left[S^{m}_{l}Z^{m}_{l\,\,\,\,cb}+\check{S}^{m}_{l}\epsilon_{cb}Z^{m}_{l}+\frac{1}{r_g^2}s^{m}_{l}X^{m}_{l\,\,\,\,cb}\right]~.\label{Eq:PerturbationSab}
\end{align}
Note that, in general, neither $\mathbb{S}$ nor its perturbations are symmetric. This is why the harmonic coefficients $S^{mA}_{l}$ and $\tilde{S}^{m}_{l\,\,\,B}$, and also $s^{mA}_{l}$ and $\tilde{s}^{\,m}_{\,l\,\,\,A}$, are in principle different, and $\check{S}^{m}_{l}$ is in general nonvanishing. In the following, we remove the harmonic labels $l$ and $m$ to make the notation lighter.

\subsubsection{The expression of $\Delta[\mathbb{S}^{\mu}_{\,\,\,\nu}]$
in terms of metric perturbations}\label{Sec:decomposeDeltaS}

As explained in Sec.~\ref{Sec:ReviewBimetric}, in order to obtain the perturbed components of the matrix $\mathbb{S}$ in terms of the perturbations of $g$ and $f$, we need to solve Eq.~\eqref{Eq:PertS}. Projecting this equation on the two-sphere, the scalar components of $\Delta[\mathbb{S}^{\mu}_{\,\,\,\nu}]$ can be solved explicitly, and read as
\begin{align}
  \tilde{S} & =\frac{1}{2\omega}\left(K^{(f)}-\omega^2K^{(g)}\right)~,\\
  S & =\frac{1}{2\omega}\left(G^{(f)}-\omega^2G^{(g)}\right)~,\\
  \check{S} & =0~,\\
  s & =\frac{1}{2\omega}\left(h^{(f)}-\omega^2h^{(g)}\right)~.
\end{align}
However, this is no longer the case for the vector and tensor harmonic components. In general, we have
\begin{align}
    s^{A} & =\left(\mathbb{M}^{-1}\right)^A_{\,\,\,C}g^{CB}\left(h^{(f)}_{B}-\omega^2h^{(g)}_{B}\right)~,\\
   S^{A} & =\left(\mathbb{M}^{-1}\right)^A_{\,\,\,C}g^{CB}\left(H^{(f)}_{B}-\omega^2H^{(g)}_{B}\right)~,
\end{align}
where $(\mathbb{M}^{-1})^A_{\,\,\,B}$ is the inverse of the matrix $\mathbb{M}^A_{\,\,\,B}=\omega\delta^A_{\,\,\,B}+\mathbb{S}^A_{\,\,\,B}$. Note that such inverse is well defined because $\mathbb{S}^A_{\,\,\,B}$ cannot have real negative eigenvalues \cite{Hassan:2017ugh}. Analogously, we obtain
\begin{align}
    \tilde{s}_{A} & =\left(h^{(f)}_{B}-h^{(g)}_{E}g^{ED}f_{DB}\right)\left(\mathbb{M}^{-1}\right)^{B}_{\,\,A}~,\\
     \tilde{S}_{A} & =\left(H^{(f)}_{B}-H^{(g)}_{E}g^{ED}f_{DB}\right)\left(\mathbb{M}^{-1}\right)^{B}_{\,\,A}~,
\end{align}
whereas, for the two-tensor $S^A_{\,\,\,B}$, we can just write
\begin{equation}
    \left(\mathbb{S}^A_{\,\,\,C}\delta^C_{\,\,\,D}\delta^E_{\,\,\,B}+\mathbb{S}^C_{\,\,\,B}\delta^A_{\,\,\,D}\delta^E_{\,\,\,C}\right)S^{D}_{\,\,\,E}=g^{AC}\left(H^{(f)}_{CB}-H^{(g)}_{CD}g^{DE}f_{EB}\right)~.
\end{equation}

\section{Dynamics of linear perturbations on a general spherically symmetric background}\label{Sec:First-orderEOM}

\subsection{Gauge and physical degrees of freedom}

At linear level, the perturbative gauge freedom can be parametrized in a vector field $\xi^\mu$, which defines the gauge transformation of the perturbation $\Delta[T]$ of any background tensor field $T$ as
\begin{align}
 \overline{\Delta[T]}=\Delta{T}+{\cal L}_\xi T.
\end{align}
In this sense, for any vector field $\xi^\mu$, $\overline{\Delta[T]}$ and $\Delta[T]$ physically represent the same perturbation of $T$. If we perform the corresponding harmonic decomposition of the vector field,
\begin{equation}
    \xi_{\mu}\, dx^\mu=\sum_{l=0}^\infty\sum _{m=-l}^l \Xi^{m}_{l\,\,A}Z^m_l dx^A
    +\sum_{l=1}^\infty\sum _{m=-l}^l\left[\Xi^{m}_l Z^{m}_{l\,\,\,\,a}+\xi^{m}_{l}X^m_{l\,\,\,\,a}\right] dx^a~,
\end{equation}
we observe that, for $l\geq 1$, there are three polar $\{\Xi^{m}_{l\,\,A},\,\Xi^{m}_l\}$ and one axial $\{\xi^{m}_{l}\}$ gauge degrees of freedom, while, for $l=0$, there are only two polar components encoded in $\Xi^{0}_{0\,\,A}$.

Applying the above transformation to the metric perturbations $h^{(i)}_{\mu\nu}$, 
\begin{align}
    \overline h{}^{(g)}_{\mu\nu}=h^{(g)}_{\mu\nu}+{\cal L}_\xi g_{\mu\nu}~,\\
    \overline h{}^{(f)}_{\mu\nu}=h^{(f)}_{\mu\nu}+{\cal L}_\xi f_{\mu\nu}~,
 \end{align}
it is straightforward to obtain the gauge transformation of the different harmonic coefficients,
\begin{subequations}\label{gauge_all}
\begin{align}\label{gauge11}
{\overline{H}}{}^{(i)}_{AB}&={H}^{(i)}_{AB}+\overset{(i)}{\nabla}{}_{\hspace{-0.07cm}B}{\Xi}_{A}+\overset{(i)}{\nabla}{}_{\hspace{-0.07cm}A}{\Xi}_{B}~, \quad l\geq 0~,\\\label{gauge12}
{\overline{H}}{}^{(i)}_{A}&={H}^{(i)}_{A}+{\Xi}_{A}-2v^{(i)}_A\Xi+\overset{(i)}{\nabla}{}_{\hspace{-0.07cm}A}{\Xi}~, \quad l\geq 1~,\\\label{gauge13}
{\overline{K}}{}^{(i)}&={K}^{(i)}+2v^{(i)A}{\Xi}_A-\frac{l(l+1)}{r_i^2}{\Xi}~, \quad l\geq 0~,\\\label{gauge14}
{\overline{G}}{}^{(i)}&={G}^{(i)}+\frac{2}{r_i^2}{\Xi}~, \quad l\geq 2~,\\\label{gauge15}
{\overline{h}}{}^{(i)}_{A}&={h}^{(i)}_{A}-2v^{(i)}_A\xi+\overset{(i)}{\nabla}{}_{\hspace{-0.07cm}A}{\xi}~, \quad l\geq 1~,\\\label{gauge16}
{\overline{h}}{}^{(i)}&={h}^{(i)}+2{\xi}~, \quad l\geq 2~,
\end{align}
\end{subequations}
where the barred objects are the harmonic coefficients of $\overline{h}^{(i)}_{\mu\nu}$. Note, in particular, that the transformation \eqref{gauge13} is defined for all $l\geq 0$, but, for $l=0$, one should understand $\Xi=0$.
For $l\geq2$ a standard choice in GR, where only one metric is present, say $g_{\mu\nu}$, is the Regge-Wheeler gauge, which corresponds to ${\overline{H}}{}^{(g)}_{A}=0$, $\overline{G}{}^{(g)}=0$, and $\overline{h}{}^{(g)}=0$.
One can also construct gauge-invariant variables associated to this particular gauge \cite{Brizuela:2007zza}: these are defined by Eqs.~\eqref{gauge_all} with $\xi=-h^{(g)}/2$, $\Xi=-r_g^2G^{(g)}/2$, and $\Xi_A=-H^{(g)}_A+G^{(g)}_{|A}/2$, which are the components of the generator of the infinitesimal transformation from a generic gauge to the Regge-Wheeler one.
However, in the vacuum bimetric theory there are twice as many perturbative variables as in GR, but the same amount of gauge degrees of freedom. Therefore, it is not possible to simultaneously choose the Regge-Wheeler gauge for the perturbations of both metrics. In fact, since it is not clear \textit{a priori} what gauge might be the most convenient one for the different applications of the formalism, we will refrain from imposing a specific gauge choice at the outset and we will present the equations of motion for any generic gauge.

Concerning the number of physical propagating degrees of freedom, one needs to take into account that, for $l\geq 2$, there are 12 first-class constraints in the theory, which can be classified in 3 four-vectors and thus contain 9 polar and 3 axial components. The two rank-two tensors $h_{\mu\nu}^{(f)}$ and $h_{\mu\nu}^{(g)}$ have a total of 20 (14 polar and 6 axial) components. Each first-class constraint removes 1 degree of freedom.
In addition, the pair of second-class constraints characteristic of bimetric theory kills the scalar (Boulware-Deser) ghost, which is polar.
In this way, for $l\geq 2$, the theory contains 7 (4 polar and 3 axial) physical propagating degrees of freedom.

Lower values of $l$ need separate consideration. Since tensor spherical harmonics are not defined for $l=1$, the above numbers differ, and, while the amount of first-class constraints remains as for $l\geq 2$ (9 polar and 3 axial), there are only 16 (12 polar and 4 axial) metric perturbations. Therefore, for $l=1$, there are 2 polar and 1 axial propagating degrees of freedom. Finally, for $l=0$, neither tensor nor vector harmonics are defined and there are no axial degrees of freedom. In this case, the 6 polar first-class constraints and a couple of second-class constraints remove 7 degrees of freedom from the 8 possible, which leaves just 1 propagating physical degree of freedom.

\subsection{Perturbative equations of motion}

As commented above, except for matter couplings, the action of the theory is invariant under the transformation \eqref{Eq:Exchangeg-f}. Making use of such a symmetry, it is straightforward to obtain the quantities associated to one metric from the quantities associated to the other. In this perturbative setup, it is clear how to implement  \eqref{Eq:Exchangeg-f} both for background objects and the harmonic coefficients of metric perturbations $h_{\mu\nu}^{(i)}$. Concerning the interaction terms, since \eqref{Eq:Exchangeg-f} maps the matrix $\mathbb{S}$ to its inverse $\mathbb{S}^{-1}$, the perturbations of $\mathbb{S}$ will be mapped to those of $\mathbb{S}^{-1}$. From perturbing the relation $\tilde{\mathbb{S}}^\mu{}_\nu(\tilde{\mathbb{S}}^{-1})^\nu{}_\alpha=\delta^\mu{}_\alpha$, one obtains
\begin{equation}
\Delta[(\mathbb{S}^{-1})^{\mu}_{\,\,\,\nu}]=-(\mathbb{S}^{-1})^{\mu}{}_{\alpha}\, \Delta[\mathbb{S}^{\alpha}_{\,\,\,\rho}]\,(\mathbb{S}^{-1})^{\rho}{}_{\nu}~,
\end{equation}
and from this expression one can read the harmonic coefficients of $\Delta[(\mathbb{S}^{-1})^{\mu}_{\,\,\,\nu}]$ from those of $\Delta[\mathbb{S}^{\mu}_{\,\,\,\nu}]$.

In this section we will explicitly provide the equations of motion for the perturbations of the metric $g_{\mu\nu}$, whereas the equations for the $f$-sector can be readily obtained using the symmetry \eqref{Eq:Exchangeg-f} discussed above and removing matter variables. More precisely, apart from obvious changes in the labels $g\to f$, in order to obtain the equations for the perturbations of $f_{\mu\nu}$, one should perform the changes,
\begin{equation}\label{Eq:Rules-f-g}
g_{AB} \rightarrow f_{AB}, \quad \beta_n \rightarrow \beta_{4-n},\quad m^2 \rightarrow m^2/\alpha^2,
\quad \mathbb{S}^A{}_B \rightarrow (\mathbb{S}^{-1})^A{}_B~, \quad \omega \rightarrow \omega^{-1},
\quad {\cal D}\rightarrow{\cal D}^{-1}~, \quad Q_m \rightarrow 0~,
\end{equation}
of background objects, while the harmonic coefficients of $\Delta[S^\mu{}_\nu]$ must be changed as follows,
\begin{equation}\label{Eq:StoInvSharmonics}
\begin{split}
    S^{A}_{\,\,\,B} & \to -(\mathbb{S}^{-1})^{A}_{\,\,\,C}\,S^{C}_{\,\,\,D}(\mathbb{S}^{-1})^{D}_{\,\,\,B}~,\\
    S^{A}&\to -\frac{1}{\omega}(\mathbb{S}^{-1})^{A}_{\,\,\,B}S^{B}~,\\
     \tilde{S}^{A} & \to -\omega\tilde{S}_{B}(\mathbb{S}^{-1})^{B}_{\,\,\,A}~,\\
    s^{A}&\to -\frac{1}{\omega}(\mathbb{S}^{-1})^{A}_{\,\,\,B}s^{B}~,\\
 \tilde{s}_{A} & \to -\omega\tilde{s}_{B}(\mathbb{S}^{-1})^{B}_{\,\,\,A}~,\\
     \tilde{S} & \to -\frac{1}{\omega^2}\tilde{S}~,\\
     S&\to -\frac{1}{\omega^2}S~,\\
     s&\to -s~.
\end{split}
\end{equation}
In addition, since we are assuming matter coupled only to the $g$-sector,
the perturbations of the matter stress-energy tensor $\Delta[\mathcal{T}_{\mu\nu}]$
must be taken to be identically zero to reproduce the equations for the $f$-sector,
\begin{equation}\label{eq.t0}
\Psi_{AB}\to 0~,\quad\Psi_{A}\to 0~,\quad\psi_{A}\to 0~,\quad 
\Psi\to 0~,\quad\widetilde{\Psi}\to 0~,\quad\psi\to 0~.
\end{equation}

The rest of the section is divided in two subsections where we analyze the axial (Sec.~\ref{SubSec:AxialEq}) and polar (Sec.~\ref{SubSec:PolarEq}) sectors separately. We recall that the differential part of the perturbative equations \eqref{Eq:BigravityEom_Perturbed} corresponds to the usual first-order perturbed Einstein tensor in spherical symmetry, whereas bimetric effects are encoded in the linearized effective stress-energy tensor. 

\subsubsection{Axial sector}\label{SubSec:AxialEq}

For $l=0$, all axial tensor spherical harmonics are identically zero, and the axial equations of motion are trivial. For $l\geq 1$, the axial part of the $(Ab)$ component of Eq. \eqref{Eq:BigravityEom_Perturbed}, that is the equation for $\Delta[G^{(i)}_{Ab}]$, gives
\begin{equation}\label{Eq:Axial-Ab}
\begin{split}
    &\overset{(i)}{\nabla}{}^B\overset{(i)}{\nabla}{}_{\hspace{-0.07cm}A}h^{(i)}_{B}-\overset{(i)}{\nabla}{}^B\overset{(i)}{\nabla}{}_{\hspace{-0.07cm}B} h^{(i)}_{A}-2(\overset{(i)}{\nabla}{}^Bh^{(i)}_{B})v_A^{(i)}+2(\overset{(i)}{\nabla}{}_{\hspace{-0.07cm}A}h^{(i)}_{B})v^{(i)B}-2h^{(i)}_{B}\overset{(i)}{\nabla}{}^Bv_A^{(i)} -4h^{(i)}_{B}v_A^{(i)}v^{(i)B}\\
    & -2h^{(i)}_{A}\Big(\frac{R^{(i)}}{2}-V_l^{(i)}\Big)+\frac{(l-1)(l+2)}{r_i^2}\Big(h^{(i)}v_A^{(i)}-\frac{1}{2}\overset{(i)}{\nabla}{}_{\hspace{-0.07cm}A}h^{(i)}\Big)=16\pi t^{(i)}_{A}~,
\end{split}
\end{equation}
where the potential $V^{(i)}_l$ reads as
\begin{equation}
    V^{(i)}_l\coloneqq -\frac{1}{r_i^2}+2\overset{(i)}{\nabla}{}^Av_A^{(i)}+3v_A^{(i)}v^{(i)A}+\frac{l(l+1)}{2r_i^2}~.
\end{equation}
The source term for the $g$-sector is
\begin{equation}\label{Eq:Pert-tAaxial}
    \begin{split}
        t^{(g)}_{A}  = & -\frac{m^2}{16\pi}\Big\{ \big[\beta_0+2\omega\beta_1+\omega^2\beta_2+(\beta_1+2\omega\beta_2+\omega^2\beta_3)\mathbb{S}^B_{\,\,\,B}+(\beta_2+2\omega\beta_3)\mathcal{D}\big]h^{(g)}_{A}-\big[\big(\beta_1+2\omega\beta_2\\
        & +\omega^2\beta_3+(\beta_2+2\omega\beta_3)\mathbb{S}^D_{\,\,\,D}+\beta_3\mathcal{D}\big)\delta^B_{\,\,\,C}
        -(\beta_2+2\omega\beta_3+\beta_3\mathbb{S}^F_{\,\,\,F})\mathbb{S}^B_{\,\,\,C}+\beta_3\mathbb{S}^B_{\,\,\,E}\mathbb{S}^E_{\,\,\,C}\big]\mathbb{S}^C_{\,\,\,A}h^{(g)}_{B}\\
       & -\big[\big(\beta_1+\omega\beta_2+(\beta_2+\omega\beta_3)\mathbb{S}^D_{\,\,\,D}+\beta_3\mathcal{D}\big)\delta^C_{\,\,\,A} -\big(\beta_2+\omega\beta_3+\beta_3\mathbb{S}^D_{\,\,\,D}\big)\mathbb{S}^C_{\,\,\,A}+\beta_3\mathbb{S}^C_{\,\,\,D}\mathbb{S}^D_{\,\,\,A}\big]\tilde{s}_{C}\\
    & -\big[\big(\beta_1+\omega\beta_2+(\beta_2+\omega\beta_3)\mathbb{S}^D_{\,\,\,D}+\beta_3\mathcal{D}\big)\delta^C_{\,\,\,B} -\big(\beta_2+\omega\beta_3+\beta_3\mathbb{S}^D_{\,\,\,D}\big)\mathbb{S}^C_{\,\,\,B}+\beta_3\mathbb{S}^C_{\,\,\,D}\mathbb{S}^D_{\,\,\,B}\big]g_{AC}s^{B}\Big\}\\
    &+\frac{1}{4}(Q_g-Q_m)h^{(g)}_{A}+\frac{1}{8\pi M_g^2}\psi_{A}~,
    \end{split}
\end{equation}
while the source $t_A^{(f)}$ can be obtained directly applying the rules \eqref{Eq:Rules-f-g}--\eqref{eq.t0} to the expression \eqref{Eq:Pert-tAaxial}.

For $l=1$, the axial part of the equation for $\Delta[G^{(i)}_{ab}]$ is not defined, whereas for $l\geq 2$, using the background equation \eqref{Eq:Background-eom}, it reads as
\begin{equation}\label{Eq:Axial-ab}
   2 \overset{(i)}{\nabla}{}^Ah^{(i)}_{A}-\overset{(i)}{\nabla}{}^A\overset{(i)}{\nabla}{}_{\hspace{-0.07cm}A}h^{(i)}+2\overset{(i)}{\nabla}{}_{\hspace{-0.07cm}A}\left(h^{(i)}v^{(i)A}\right)=16\pi\left(t^{(i)}-\frac{Q_i}{2}h^{(i)}\right)~,
\end{equation}
with
\begin{equation}\label{Eq:Pert-tscalar1}
     t^{(g)}  = \frac{1}{4}(Q_g-Q_m) h^{(g)}+\frac{m^2}{16\pi}\left[\beta_1 +\beta_2 \mathbb{S}^A_{\,\,\,A}+\beta_3 \mathcal{D} \right]s+\frac{1}{8\pi M_g^2}\psi~,
\end{equation}
and $t^{(f)}$ can be derived using \eqref{Eq:Rules-f-g}--\eqref{eq.t0}.

Therefore, the evolution of the axial sector is completely determined by Eqs. \eqref{Eq:Axial-Ab} and \eqref{Eq:Axial-ab}. As commented in the previous section, there is one gauge degree of freedom, which one can fix. With the equations at hand, we can analyze more explicitly the number of propagating degrees of freedom in this sector. For $l=1$, there are four equations, all of them contained in the relation \eqref{Eq:Axial-Ab}. Making explicit the second-order derivative terms and expanding in a generic chart $x^A=(x^0,x^1)$, they can be combined to give, schematically 
\begin{subequations}\label{eq.schematicaxial}
   \begin{equation}\label{eq.axial1}
    \blacksquare \partial_0\partial_1 h^{(i)}_{1}- \blacksquare \partial_1\partial_1 h^{(i)}_{0}=\dots,
\end{equation}
\begin{equation}\label{eq.axial2}
    \blacksquare \partial_0\partial_1 h^{(i)}_{0}- \blacksquare \partial_0\partial_0 h^{(i)}_{1}=\dots~,
\end{equation} 
\end{subequations}
for $i=\{f,g\}$, where $\blacksquare$ stands for background terms, while the dots encode first-order derivatives and terms with no derivatives. Since there are no second-order time derivatives, Eqs. \eqref{eq.axial1} are constraint equations, while Eqs. \eqref{eq.axial2} can be understood as evolution equations for the two functions $h_1^{(f)}$ and $h_1^{(g)}$. However, the remaining axial gauge degree of freedom kills one of those, for instance by choosing $h_1^{(f)}=0$, which leaves one single propagating axial degree of freedom for $l=1$.

Now, for $l\geq 2$, in addition to the four equations \eqref{eq.schematicaxial}, one also has
\eqref{Eq:Axial-ab}, with principal part
\begin{equation}
    \blacksquare \partial_0\partial_0 h^{(i)}+ \blacksquare \partial_1\partial_1 h^{(i)}=\dots~.
\end{equation}
These can be understood as two evolution equations for $h^{(g)}$ and $h^{(f)}$. There is the same amount of gauge freedom as for $l=1$, and thus one ends up with three propagating axial degrees of freedom for $l\geq 2$.

Nonetheless, it is highly nontrivial to obtain the corresponding master variables that would obey unconstrained hyperbolic equations and would thus encode complete physical information on the problem. Following the procedure presented by Gerlach-Sengupta \cite{PhysRevD.19.2268}, one can define the following scalar functions\footnote{In fact, here we are using a rescaled variable since it leads to a simpler form of the evolution equation. The variable introduced by Gerlach-Sengupta reads as $\Pi^{(i)}_{\rm GS}=\Pi^{(i)}/r_i^3$.}
\begin{equation}\label{Eq:DefPi}
    \Pi^{(i)}=r_i^3\epsilon^{AB}\overset{(i)}{\nabla}{}_{\hspace{-0.07cm}B}\left(r_i^{-2}h^{(i)}_{A}\right)~,
\end{equation}
so that, taking then the curl of Eq. \eqref{Eq:Axial-Ab}, yields
\begin{equation}\label{Eq:wavePi}
\begin{split}
  \overset{(i)}{\nabla}{}^A\overset{(i)}{\nabla}{}_{\hspace{-0.07cm}A}\Pi^{(i)}- \widetilde{V}{}^{(i)}\Pi^{(i)}+ & \frac{(l-1)(l+2)}{2}r_i^3\epsilon^{AB}\overset{(i)}{\nabla}{}_{\hspace{-0.07cm}B}\left(h^{(i)}v_A^{(i)}-\frac{1}{2}\overset{(i)}{\nabla}{}_{\hspace{-0.07cm}A}h^{(i)}\right) =8\pi r_i^3\epsilon^{AB}\overset{(i)}{\nabla}{}_{\hspace{-0.07cm}B}\left(t^{(i)}_{A}-\frac{Q_i}{2}h^{(i)}_{A}\right)~,  
\end{split}
\end{equation}
for $l\geq 1$, where 
\begin{equation}
    \widetilde{V}^{(i)}=\frac{l(l+1)-3}{r_i^2}+3v^{(i)A}v^{(i)}_A~.
\end{equation}
Since in GR there is only one copy of equation \eqref{Eq:wavePi}, say for $i=g$, introducing the new matter invariant $\phi_A=t^{(g)}_{A}-\frac{Q_g}{2}h^{(g)}_{A}$ \cite{PhysRevD.19.2268}, one can use the remaining gauge freedom to set $h^{(g)}=0$ (for $l\geq 2$). In this way, in GR this equation is uncoupled to the rest of the metric perturbations and thus $\Pi^{(g)}$ follows an unconstrained evolution equation, which, for vacuum, reduces to the Regge-Wheeler equation \cite{Regge:1957td}. However, in bimetric gravity there is not enough gauge freedom to set both $h^{(i)}$ to zero and, in addition, the sources $t_A^{(i)}$ do not only correspond to matter perturbations, but they are complicated functions (cf.~Eq.~\eqref{Eq:Pert-tAaxial}) of the metric perturbations. Therefore, the variables $\Pi^{(i)}$ defined by \eqref{Eq:DefPi} do not obey unconstrained master equations uncoupled to other metric perturbations.

\subsubsection{Polar sector}\label{SubSec:PolarEq}
In this subsection, we provide the set of equations for the polar perturbations of the $g$-sector, while the equations corresponding to the $f$-sector can be obtained by applying the rules \eqref{Eq:Rules-f-g}--\eqref{eq.t0}. On the one hand, the equation for $\Delta[G^{(i)}_{AB}]$ gives, for $l\geq 0$,
\begin{equation}\label{Eq:Polar-AB}
    \begin{split}
       & 2\Big(\overset{(i)}{\nabla}{}_{\hspace{-0.07cm}B}H^{(i)}_{CA}+\overset{(i)}{\nabla}{}_{\hspace{-0.07cm}A}H^{(i)}_{CB}-2\mathfrak{g}^{(i)}_{AB}\overset{(i)}{\nabla}{}^DH^{(i)}_{DC}+\mathfrak{g}^{(i)}_{AB}\overset{(i)}{\nabla}{}_{\hspace{-0.07cm}C}H^{(i)D}_{D}\Big)v^{(i)C}+2H^{(i)}_{AB}V_l^{(i)}-\mathfrak{g}^{(i)}_{AB}\Big[\frac{l(l+1)}{r_i^2}H^{(i)C}_{C}\\
       & +6H^{(i)}_{DC}v^{(i)D}v^{(i)C}+4H^{(i)}_{DC}\overset{(i)}{\nabla}{}^Cv^{(i)D}\Big]-\frac{l(l+1)}{r_i^2}\Big[\overset{(i)}{\nabla}{}_{\hspace{-0.07cm}B}H^{(i)}_{A}+\overset{(i)}{\nabla}{}_{\hspace{-0.07cm}A}H^{(i)}_{B}-2\mathfrak{g}^{(i)}_{AB}(\overset{(i)}{\nabla}{}^CH^{(i)}_{C}+H^{(i)}_{C}v^{(i)C})\Big]\\
       &+ \mathfrak{g}^{(i)}_{AB}\Big[6\overset{(i)}{\nabla}{}_{\hspace{-0.07cm}C}K^{(i)}v^{(i)C}+2\overset{(i)}{\nabla}{}^C\overset{(i)}{\nabla}{}_{\hspace{-0.07cm}C}K^{(i)}-\frac{(l-1)(l+2)}{r_i^2}\Big(K^{(i)}+\frac{l(l+1)}{2}G^{(i)}\Big)\Big]\\
       & -2\Big(\overset{(i)}{\nabla}{}_{\hspace{-0.07cm}B}\overset{(i)}{\nabla}{}_{\hspace{-0.07cm}A}K^{(i)}+\overset{(i)}{\nabla}{}_{\hspace{-0.07cm}A}K^{(i)}v^{(i)}_{B}+\overset{(i)}{\nabla}{}_{\hspace{-0.07cm}B}K^{(i)}v^{(i)}_{A}\Big)=16\pi T^{(i)}_{AB}~,
    \end{split}
\end{equation}
with
\begin{equation}\label{Eq:Pert-tAB}
    \begin{split}
        T^{(g)}_{AB}= &  -\frac{m^2}{16\pi}\Big\{ \Big[(2\beta_0+4\omega\beta_1+2\omega^2\beta_2)+(2\beta_1+4\omega\beta_2+2\omega^2\beta_3)\mathbb{S}^C_{\,\,\,C}+(2\beta_2+4\omega\beta_3)\mathcal{D}\Big]H^{(g)}_{AB}\\
        &-2\Big[\left(\beta_1+2\omega\beta_2+\omega^2\beta_3+(\beta_2+2\omega\beta_3)\mathbb{S}^D_{\,\,\,D}+\beta_3\mathcal{D}\right)\mathbb{S}^C_{\,\,(A}\delta^E_{\,\,\,B)} -\big((\beta_2+2\omega\beta_3+\beta_3\mathbb{S}^F_{\,\,\,F})\mathbb{S}^C_{\,\,\,D}\\
        & -\beta_3\mathbb{S}^F_{\,\,\,D}\mathbb{S}^C_{\,\,\,F}\big) \mathbb{S}^D_{\,\,(A}\delta^E_{\,\,\,B)} \Big]H^{(g)}_{EC} -2\Big[\left(\beta_1+2\omega\beta_2+\omega^2\beta_3+(\beta_2+2\omega\beta_3)\mathbb{S}^D_{\,\,\,D}+\beta_3\mathcal{D}\right)\delta^E_{\,\,\,C}\\
        & -(\beta_2+2\omega\beta_3+\beta_3\mathbb{S}^F_{\,\,\,F})\mathbb{S}^E_{\,\,\,C}+\beta_3\mathbb{S}^E_{\,\,\,D}\mathbb{S}^D_{\,\,\,C}\Big]g_{E(B}S^C_{\,\,A)} +2\Big[\big(\beta_1+2\omega\beta_2+\omega^2\beta_3\\
        & +(\beta_2+2\omega\beta_3)\mathbb{S}^D_{\,\,\,D}+\beta_3\mathcal{D}\big)g_{AB}-\left((\beta_2+2\omega\beta_3+\beta_3\mathbb{S}^F_{\,\,\,F})\delta^E_{\,\,\,D}-\beta_3\mathbb{S}^E_{\,\,\,D}\right)g_{E(B}\mathbb{S}^D_{\,\,\,A)}\Big]S^C_{\,\,\,C}\\
        & +2\Big[\left((\beta_2+2\omega\beta_3+\beta_3\mathbb{S}^F_{\,\,\,F})\delta^D_{\,\,\,E}-\beta_3\mathbb{S}^D_{\,\,\,E}\right)g_{C(B}\mathbb{S}^E_{\,\,\,A)}+\beta_3g_{E(B}\mathbb{S}^E_{\,\,\,A)}\mathbb{S}^D_{\,\,\,C}-\beta_3g_{E(B}\mathbb{S}^D_{\,\,\,A)}\mathbb{S}^E_{\,\,\,C}\\
        & -\left((\beta_2+2\omega\beta_3+\beta_3\mathbb{S}^F_{\,\,\,F})\mathbb{S}^D_{\,\,\,C}-\beta_3\mathbb{S}^D_{\,\,\,E}\mathbb{S}^E_{\,\,\,C}\right)g_{AB}\Big]S^C_{\,\,\,D} +4\Big[(\beta_1+\omega\beta_2+(\beta_2+\omega\beta_3)\mathbb{S}^C_{\,\,\,C}\\
        & +\beta_3\mathcal{D})g_{AB}-((\beta_2+\omega\beta_3+\beta_3\mathbb{S}^D_{\,\,\,D})\delta^E_{\,\,\,C}-\beta_3\mathbb{S}^E_{\,\,\,C})g_{E(B}\mathbb{S}^C_{\,\,A)}\Big]\tilde{S}\Big\}+\frac{1}{8\pi M_g^2}\Psi_{AB}~.
\end{split}
\end{equation}
On the other hand, from the equation for $\Delta[G^{(i)}_{Ab}]$, and for $l\geq 1$, one obtains
\begin{equation}\label{Eq:Polar-Ab}
    \begin{split}
      & \overset{(i)}{\nabla}{}^BH^{(i)}_{AB}-\overset{(i)}{\nabla}{}_{\hspace{-0.07cm}A}H^{(i)B}_{B}+H^{(i)B}_{B}v^{(i)}_A+\overset{(i)}{\nabla}{}^B\overset{(i)}{\nabla}{}_{\hspace{-0.07cm}A}H^{(g)}_B-\overset{(i)}{\nabla}{}^B\overset{(i)}{\nabla}{}_{\hspace{-0.07cm}B}H^{(i)}_{A}-2v^{(i)}_A\overset{(i)}{\nabla}{}^BH^{(i)}_{B} +2v^{(i)B}\overset{(i)}{\nabla}{}_{\hspace{-0.07cm}A}H^{(i)}_{B}\\
    &-2H^{(i)}_{B}\overset{(i)}{\nabla}{}^Bv^{(i)}_A-4H^{(i)}_{B}v^{(i)B}v^{(i)}_A-2H^{(i)}_{A}\Big(\frac{R^{(i)}}{2}-V^{(i)}_0\Big)-\overset{(i)}{\nabla}{}_{\hspace{-0.07cm}A}K^{(i)}-\frac{(l-1)(l+2)}{2}\overset{(i)}{\nabla}{}_{\hspace{-0.07cm}A}G^{(i)}=16\pi T^{(i)}_{A}~,
    \end{split}
\end{equation}
with 
\begin{equation}\label{Eq:Pert-tApolar}
    \begin{split}
        T^{(g)}_{A}  = & -\frac{m^2}{16\pi}\Big\{ \big[\beta_0+2\omega\beta_1+\omega^2\beta_2+(\beta_1+2\omega\beta_2+\omega^2\beta_3)\mathbb{S}^B_{\,\,\,B}+(\beta_2+2\omega\beta_3)\mathcal{D}\big]H^{(g)}_{A}-\big[\big(\beta_1+2\omega\beta_2\\
        & +\omega^2\beta_3+(\beta_2+2\omega\beta_3)\mathbb{S}^D_{\,\,\,D}+\beta_3\mathcal{D}\big)\delta^C_{\,\,\,A}-(\beta_2+2\omega\beta_3+\beta_3\mathbb{S}^{D}_{\,\,\,D})\mathbb{S}^C_{\,\,\,A}+\beta_3\mathbb{S}^C_{\,\,\,D}\mathbb{S}^{D}_{\,\,\,A})\big]\mathbb{S}^B_{\,\,\,C}H^{(g)}_{B}\\
       & -\big[\big(\beta_1+\omega\beta_2+(\beta_2+\omega\beta_3)\mathbb{S}^D_{\,\,\,D}+\beta_3\mathcal{D}\big)\delta^B_{\,\,\,A} -\big(\beta_2+\omega\beta_3+\beta_3\mathbb{S}^D_{\,\,\,D}\big)\mathbb{S}^B_{\,\,\,A}+\beta_3\mathbb{S}^B_{\,\,\,D}\mathbb{S}^D_{\,\,\,A}\big]\tilde{S}_{B}\\
    & -\big[\big(\beta_1+\omega\beta_2+(\beta_2+\omega\beta_3)\mathbb{S}^D_{\,\,\,D}+\beta_3\mathcal{D}\big)\delta^C_{\,\,\,B} -\big(\beta_2+\omega\beta_3+\beta_3\mathbb{S}^D_{\,\,\,D}\big)\mathbb{S}^C_{\,\,\,B}+\beta_3\mathbb{S}^C_{\,\,\,D}\mathbb{S}^D_{\,\,\,B}\big]g_{AC}S^{B}\Big\}\\
    &+\frac{1}{4}(Q_g-Q_m)H^{(g)}_{A}+\frac{1}{8\pi M_g^2}\Psi_{A}~.
    \end{split}
\end{equation}
Finally, $\Delta[G^{(i)}_{ab}]$ gives, for $l\geq 2$,
\begin{equation}\label{Eq:Polar-ab-1}
    \begin{split}
        -H^{(i)A}_{A}+2\overset{(i)}{\nabla}{}^AH^{(i)}_{A}-r_i^2\overset{(i)}{\nabla}{}^A\overset{(i)}{\nabla}{}_{\hspace{-0.07cm}A}G^{(i)}-2r_i^2\overset{(i)}{\nabla}{}_{\hspace{-0.07cm}A}G^{(i)}v^{(i)A}=16\pi\Big(T^{(i)}-\frac{r_i^2 Q_i}{2}G^{(i)}\Big)~,
    \end{split}
\end{equation}
with
\begin{equation}\label{Eq:Pert-tscalar2}
     T^{(g)}  = \frac{1}{4}(Q_g-Q_m) r_g^2G^{(g)}+\frac{m^2r_g^2}{16\pi}\left[\beta_1 +\beta_2 \mathbb{S}^A_{\,\,\,A}+\beta_3 \mathcal{D}\right]S+\frac{1}{8\pi M_g^2}\Psi~,
\end{equation}
and, for $l\geq 0$,
\begin{equation}\label{Eq:Polar-ab-2}
    \begin{split}
       & -\overset{(i)}{\nabla}{}_{\hspace{-0.07cm}B}\overset{(i)}{\nabla}{}_{\hspace{-0.07cm}A}H^{(i)AB}+\overset{(i)}{\nabla}{}^B\overset{(i)}{\nabla}{}_{\hspace{-0.07cm}B}H^{(i)A}_{A}-2\overset{(i)}{\nabla}{}_{\hspace{-0.07cm}A}H^{(i)AB}v^{(i)}_{B}+\overset{(i)}{\nabla}{}_{\hspace{-0.07cm}B}H^{(i)A}_{A}v^{(i)B}-2H^{(i)AB}(\overset{(i)}{\nabla}{}_{\hspace{-0.07cm}B}v^{(i)}_{A}+v^{(i)}_Av^{(i)}_B) \\
    & +\Big(\frac{R^{(i)}}{2}-\frac{l(l+1)}{2r_i^2}\Big)H^{(i)A}_{A}+\frac{l(l+1)}{r_i^2}\overset{(i)}{\nabla}{}^AH^{(i)}_{A}+\overset{(i)}{\nabla}{}^A\overset{(i)}{\nabla}{}_{\hspace{-0.07cm}A}K^{(i)}+2\overset{(i)}{\nabla}{}_{\hspace{-0.07cm}A}K^{(i)}v^{(i)A}=16\pi \Big(\widetilde{T}^{(i)}-\frac{Q_i}{2}K^{(i)}\Big)~,
    \end{split}
\end{equation}
where
\begin{equation}\label{Eq:Pert-tscalar3}
\begin{split}
     \widetilde{T}^{(g)}  = & -\frac{m^2}{16\pi}\Big\{\left[\beta_1 +\beta_2 \mathbb{S}^A_{\,\,\,A}+\beta_3 \mathcal{D}\right]\tilde{S}-\left[(\beta_2+\omega\beta_3+\beta_3\mathbb{S}^D_{\,\,\,D})\delta^C_{\,\,\,A}-\beta_3\mathbb{S}^C_{\,\,\,A})\right]\mathbb{S}^B_{\,\,\,C}S^A_{\,\,\,B}\\
      &+\left[\beta_1+\omega\beta_2+(\beta_2+\omega\beta_3)\mathbb{S}^B_{\,\,\,B}+\beta_3\mathcal{D}\right]S^A_{\,\,\,A}\Big\}+\frac{1}{4}(Q_g-Q_m)K^{(g)}+\frac{1}{8\pi M_g^2}\widetilde{\Psi}~.
\end{split}
\end{equation}
Again, the polar components of $\Delta[t^{(f)}_{\mu\nu}]$ can be derived using \eqref{Eq:Rules-f-g}- \eqref{eq.t0}.

The number of propagating degrees of freedom in this sector can be analyzed following
the same rationale as used in the axial sector. However, the polar case is much more involved, due to the greater number of equations and variables. Concerning master equations, we would like to note that the construction of a polar master variable for a generic background is an open question even in GR, and there are results only for certain specific backgrounds, like the Zerilli variable for vacuum \cite{PhysRevLett.24.737}.

\section{Static backgrounds}\label{Sec:StaticBackground}

Next, we proceed to apply the formalism developed in previous sections to specific backgrounds of interest. In this section we will assume that the background metric $g_{\mu\nu}$ is static, that is, it contains a hypersurface-orthogonal Killing field $\partial_t$. Since exact bidiagonal solutions have been shown to lead to instabilities \cite{Babichev:2013una, Brito:2013wya, Babichev:2014oua, Torsello:2017cmz}, such backgrounds will not be treated. Here we will focus instead exclusively on nonbidiagonal backgrounds, thus assuming that there does not exist a chart such that the metrics $f_{\mu\nu}$ and $g_{\mu\nu}$ are both diagonal. As it is well known \cite{Volkov:2012wp, Volkov2015}, imposing a staticity condition on $g_{\mu\nu}$ implies that $f_{\mu\nu}$ is also static, and has a Killing vector field $\partial_T=\frac{1}{\dot{T}}\partial_t$ that is collinear with $\partial_t$. (Here and in the following an overdot is used to denote a derivative with respect to $t$.) We exhibit the general equations of motion for perturbations around such a static nonbidiagonal background in vacuo, obtained as a particular case of the equations derived in Sec.~\ref{Sec:First-orderEOM}.  Finally, we discuss the special case where the Killing vector fields of both metrics coincide, that is for $\dot{T}=constant$.

\subsection{Nonbidiagonal background metrics with a static $g_{\mu\nu}$}

Following Ref.~\cite{Volkov2015}, let us thus begin with the most general nonbidiagonal ansatz with a static form for $g_{\mu\nu}$:
\begin{align}
    g_{\mu\nu}\de x^\mu\de x^\nu & =-U(r)\de t^2+V(r)\de r^2+r^2\left(\de \theta^2+\sin^2\theta\de \varphi ^2\right)~,\\
    f_{\mu\nu}\de x^\mu\de x^\nu & =-A(t,r)\de t^2+B(t,r)\de r^2+C(t,r)\de t\de r+r_f^2(t,r)\left(\de \theta^2+\sin^2\theta\de \varphi ^2\right)~,
\end{align}
where $r_f$ is positive, $C\neq 0$, and the chart is valid for $U\neq 0$
and $V\neq 0$.
Since $g_{\mu\nu}$ is diagonal and independent of $t$, its Einstein tensor $G^{(g)\mu}_{\hspace{0.5cm}\,\nu}$ is also diagonal. Moreover, it follows from the equations of motion \eqref{Eq:BigravityEom_matter-g} (with $\mathcal{T}_{\mu\nu}=0$) that $V^{(g)\mu}_{\hspace{0.5cm}\,\,\nu}$ must also be diagonal on solutions. This implies the following algebraic constraint
\begin{equation}
    V^{(g)t}_{\hspace{0.5cm}\,r}\propto V^{(g)r}_{\hspace{0.5cm}\,t}\propto C(r^2\beta_1+2r\beta_2r_f+\beta_3r_f^2)=0~.
\end{equation}
Since we are considering nonbidiagonal solutions with $C\neq 0$, this equation translates into the condition $r_f=\omega r$, with $\omega$ a positive root of
\begin{equation}\label{Eq:Condition-1-nonbidiagonal}
    \beta_1+2\beta_2\omega+\beta_3\omega^2=0~.
\end{equation}
Moreover, the Bianchi constraint $\nabla^{(g)\mu}V^{(g)}_{\mu\nu}=0$ implies
\begin{equation}
    (\beta_2+\omega\beta_3)\left[(\omega-\mathbb{S}^t_{\,\,t})(\omega-\mathbb{S}^r_{\,\,r})-\mathbb{S}^t_{\,\,r}\mathbb{S}^r_{\,\,t}\right]=0~.
\end{equation}
Thus, leaving aside the particular choice of parameters $(\beta_2+\omega\beta_3)=0$, the combination of terms in square brackets must vanish.\footnote{If $\beta_2+\omega\beta_3=0$, the condition \eqref{Eq:Condition-1-nonbidiagonal} implies $\beta_1+\omega\beta_2=0$. As we will see below, in this special case, the metrics decouple even at linear level, reducing the perturbation equations to the linearized Einstein equations.}
This leads us to the following relation in terms of the metric functions
\begin{equation}\label{Eq:Condition-2-nonbidiagonal}
 C^2=-4(B-\omega^2V)(A-\omega^2 U)~.
\end{equation}
Note, in particular, that the reality of the metric restricts the right-hand side of this expression to be strictly non-negative.

Next, imposing \eqref{Eq:Condition-1-nonbidiagonal} and \eqref{Eq:Condition-2-nonbidiagonal}, it can be shown that the equations of motion \eqref{Eq:BigravityEom_matter-g} for the background at hand boil down to the Einstein equations,
\begin{equation}\label{Eq:BigravityEom_nonbidiagonal-g}
 G_{\mu\nu}^{(g)}+m^2\Lambda_g g_{\mu\nu}=0~,
\end{equation}
with the effective cosmological constant $\Lambda_g\coloneqq\beta_0+2\omega\beta_1+\omega^2\beta_2$ defined in terms of the parameters of the theory. Therefore, the standard Birkhoff theorem with cosmological constant applies, and the solution for the metric coefficients is
\begin{equation}
  U=\frac{1}{V}=\Sigma_g~,\quad  \mbox{with}\quad \Sigma_g := 1-\frac{2\mu_g}{r}-\frac{m^2\Lambda_g}{3}r^2~,
\end{equation}
which completely determines $g_{\mu\nu}$ as the Schwarzschild-(anti)de Sitter metric, depending on the sign of $\Lambda_g$.

Now, under the above assumptions, the equations of motion for $f_{\mu\nu}$ \eqref{Eq:BigravityEom_matter-f} are decoupled from $g_{\mu\nu}$ and they also reduce to the Einstein equations,  
\begin{equation}\label{Eq:BigravityEom_nonbidiagonal-f}
G_{\mu\nu}^{(f)}+\frac{m^2}{\alpha^2}\Lambda_f f_{\mu\nu} =0~,
\end{equation}
with the corresponding cosmological constant given by $\Lambda_f\coloneqq\frac{1}{\omega^2}(\beta_2+2\omega\beta_3+\omega^2\beta_4)$. In addition, the metric functions must also obey the nonbidiagonal condition \eqref{Eq:Condition-2-nonbidiagonal}. In order to solve these equations, it is convenient to change to new coordinates $(T,r_f)$, with $T=T(t,r)$, where the metric $f_{\mu\nu}$ becomes diagonal,
\begin{equation}
    f_{\mu\nu}\de x^{\mu}\de x^{\nu}=-f_{TT}(T,r_f)\de T^2+f_{r_fr_f}(T,r_f)\de r_f^2+r_f^2(\de \theta^2+\sin^2\theta\de \varphi^2)~.
\end{equation}
The solution of Eq.~\eqref{Eq:BigravityEom_nonbidiagonal-f} in these new coordinates is once again the diagonal form of the Schwarzschild-(anti)de Sitter metric
\begin{equation}
    f_{\mu\nu}\de x^{\mu}\de x^{\nu}=-\Sigma_f \de T^2+\frac{1}{\Sigma_f}\de r_f^2+r_f^2(\de \theta^2+\sin^2\theta\de \varphi^2)~,
\end{equation}
with $\Sigma_f=1-\frac{2\mu_f}{r_f}-\frac{m^2\Lambda_f}{3\alpha^2}r_f^2$. Transforming back to the original $(t,r)$ coordinates, one finds the relations
\begin{equation}
    A(t,r)=\Sigma_f\Dot{T}^2~,\quad B(t,r)=-\Sigma_f T^{\prime\,2}+\Sigma_f^{-1}\omega^2~,\quad C(t,r)=-2\Sigma_f\Dot{T}T^{\prime}~,
\end{equation}
which, upon substitution into Eq.~\eqref{Eq:Condition-2-nonbidiagonal}, yield the following partial differential equation for the unknown function $T=T(t,r)$,
\begin{equation}\label{eq:Tequation}
T^{\prime\,2}=\left(\frac{1}{\Sigma_g}-\frac{1}{\Sigma_f} \right)\left(\frac{\Dot{T}^2}{\Sigma_g}-\frac{\omega^2}{\Sigma_f}\right).
\end{equation}
Here we have defined $\Dot{T}\coloneqq\partial T/\partial t$ and $T^\prime\coloneqq\partial T/\partial r$. Note that, in general, the function $T$ will depend on both $(t,r)$. In fact, for $C$ to be nonvanishing, so as to ensure a nonbidiagonal form of the metrics, neither $\dot{T}$ nor $T^\prime$ can vanish. In particular, this excludes the case where the two metrics describe black holes with the same mass and cosmological constant, since that would imply $\Sigma_g=\Sigma_f$ and thus, following \eqref{eq:Tequation}, $T^{\prime}=0$. Since \eqref{eq:Tequation} is a nonlinear partial differential equation, there is no systematic procedure to obtain its general solution $T=T(t,r)$. In addition, the reality conditions imply that the right-hand side of \eqref{eq:Tequation} must be non-negative, which, in general, will impose certain restrictions on $\dot T$ (or, if one had a general solution at hand, on the corresponding integration constants). Interestingly, in regions where $\Sigma_f\Sigma_g<0$, the right-hand side of \eqref{eq:Tequation} is positive definite, and thus $\dot T$ is unrestricted by this condition. Note also that, in terms of the function $T$, for this nonbidiagonal ansatz, the matrix $\mathbb{S}$ can be written in the following compact form,
\begin{subequations}
    \begin{align}
        \mathbb{S}^t_{\,\,t}&=\frac{\Sigma_f\Dot{T}^2+\Sigma_g\omega|\Dot{T}|}{\Sigma_g(\omega+|\Dot{T}|)}~,\\
        \mathbb{S}^r_{\,\,r}&=\frac{(\omega^2+\Dot{T}^2+\omega|\Dot{T}|)\Sigma_g-\Dot{T}^2\Sigma_f}{\Sigma_g(\omega+|\Dot{T}|)}~,\\
        \mathbb{S}^t_{\,\,r}&=-\Sigma_g^2\mathbb{S}^r_{\,\,t}=\frac{\Sigma_f T^{\prime}\Dot{T}}{\Sigma_g(\omega+|\Dot{T}|)}~.
    \end{align}
\end{subequations}
From these expressions, it is straightforward to conclude that the matrix $\mathbb{S}$ will be real as long as $T$ is real.

There is, however, one specific interesting case where Eq.~\eqref{eq:Tequation} can be solved. Namely, if one assumes that the Killing vector field of both metrics coincide, and thus $\dot{T}$ is constant, the equation can then be reduced to the quadrature,
\begin{equation}\label{Eq:solTquadrature}
 T=c\, t+\int dr\,\sqrt{\left(\frac{1}{\Sigma_g}-\frac{1}{\Sigma_f} \right)\left(\frac{c^2}{\Sigma_g}-\frac{\omega^2}{\Sigma_f}\right)},
\end{equation}
with $c$ an integration constant. Owing to the reality conditions discussed above, this integration constant is not completely free in general, and it is constrained so that the argument of the square root is positive, a condition that will depend on the specific parameters (mass and cosmological constant) of the black holes and on the range of $r$. Remarkably, the choice $c^2=\omega^2$ is the only one that reduces the argument of the square root to a perfect square, and therefore it is valid for any parameter of the black holes and any range of $r$. Furthermore, we note that the background geometry considered in Ref.~\cite{Babichev:2015zub} can be obtained as a particular case of our more general \eqref{Eq:solTquadrature} with $c=\omega$ and $\Lambda_g=\Lambda_f=0$.

At background level, the interaction between the two metric sectors only manifests itself through the cosmological constants $\Lambda_g$ and $\Lambda_f$, so that the two metrics are effectively decoupled. Therefore, one could treat both metrics as independent and take a different coordinate frame for each, for instance, such that both are diagonal (i.e.,~$(t,r)$ for $g_{\mu\nu}$ and $(T,r_f)$ for $f_{\mu\nu}$). In this sense, $c$ does not have a physical impact on the background geometry, and, in particular, no curvature invariant depends on $c$. Hence, at the background level, this constant only appears when one relates the two metrics. For instance, it affects the relative tilt of the light-cones of $g_{\mu\nu}$ and $f_{\mu\nu}$ (for an analysis of the causal structure in the general case see Refs.~\cite{Hassan:2017ugh, Kocic:2020pnm}).
However, at a perturbative level the two sectors are indeed coupled, and the constant $c$ appears in the equations of motion in a nontrivial way.

\subsection{Linear perturbations on a static nonbidiagonal background}

Here we compute the source terms in the equations of motion for linear perturbations around nonbidiagonal static backgrounds. The Killing vector field of $g_{\mu\nu}$ is $\partial_t$. Then, under these conditions, $f_{\mu\nu}$ is also static, though its Killing vector field $\partial_T$ generically does not coincide with $\partial_t$, but is instead defined in terms of the function $T=T(t,r)$ that solves Eq.~\eqref{eq:Tequation}.

For a general static nonbidiagonal spherically symmetric ansatz, the expressions for the axial harmonic components of the perturbed effective stress-energy tensor for the metric $g_{\mu\nu}$, Eqs.~\eqref{Eq:Pert-tAaxial} and \eqref{Eq:Pert-tscalar1}, imposing 
$\mathcal{T}_{\mu\nu}=\Delta[\mathcal{T}_{\mu\nu}]=0$,
take the form
\begin{align}
    &t^{(g)}_{A}=-\frac{m^2}{16\pi}\left(2\Lambda_g h^{(g)}_{A}-\frac{(\beta_1+\omega \beta_2)(h^{(f)}_{B}-\omega^2 h^{(g)}_{B})\mathbb{Q}^B_{\,\,\,\,A}}{\omega^2\Sigma_g(|\Dot{T}|+\omega)}\right)~,\\
    &t^{(g)}=-\frac{m^2 }{16\pi}\left(\Lambda_g h^{(g)}+\frac{(\beta_1+\omega \beta_2)(h^{(f)}-\omega^2 h^{(g)})(|\Dot{T}|-\omega)}{2\omega^2}\right)~,
\end{align}
where we have introduced the two-by-two matrix
\begin{equation}
    \mathbb{Q}^{A}_{\,\,\,\,B}:=\begin{pmatrix}
        \dot{T}^2(\Sigma_f-\Sigma_g) & \Sigma_f \dot{T}T^\prime\\
       \vspace{-0.2cm} &\\
        -\Sigma_f \dot{T}T^\prime \Sigma_g^2 & \omega^2\Sigma_g-\dot{T}^2\Sigma_f
    \end{pmatrix}~.
\end{equation}
At this point, it is clear that when the Killing vector fields of both metrics coincide (and therefore $\Dot{T}=c$), the constant $c$ will appear explicitly in the equations of motion through the source terms. Similarly, for the axial harmonic components of $\Delta[t_{\mu\nu}^{(f)}]$, we have
\begin{align}
    &t^{(f)}_{A}=-\frac{m^2}{16\pi\alpha^2}\left(2\Lambda_f h^{(f)}_{A}+\frac{(\beta_1+\omega \beta_2)(h^{(f)}_{B}-\omega^2 h^{(g)}_{B})\mathbb{Q}^B_{\,\,\,\,A}}{\omega^3\Sigma_g|\Dot{T}|(|\Dot{T}|+\omega)}\right)~,\\
   & t^{(f)}=-\frac{m^2 }{16\pi\alpha^2}\left(\Lambda_f h^{(f)}-\frac{(\beta_1+\omega \beta_2)(h^{(f)}-\omega^2 h^{(g)})(|\Dot{T}|-\omega)}{2\omega^3|\Dot{T}|}\right)~.
\end{align}
As for the polar components, Eqs.~\eqref{Eq:Pert-tAB}, \eqref{Eq:Pert-tApolar},\eqref{Eq:Pert-tscalar2}, and \eqref{Eq:Pert-tscalar3}, boil down to
\begin{align}
  & T^{(g)}_{AB} =-\frac{m^2}{8\pi}\left(\Lambda_g H^{(g)}_{AB}-\frac{(\beta_1+\omega \beta_2)(K^{(f)}-\omega^2 K^{(g)})\mathbb{P}_{AB}}{\omega^2(|\Dot{T}|+\omega)}\right)~,\\
  &  T^{(g)}_{A} =-\frac{m^2}{16\pi}\left(2\Lambda_g H^{(g)}_{A}-\frac{(\beta_1+\omega \beta_2)(H^{(f)}_{B}-\omega^2 H^{(g)}_{B})\mathbb{Q}^B_{\,\,\,\,A}}{\omega^2\Sigma_g(|\Dot{T}|+\omega)}\right)~,\\
    &T^{(g)}=-\frac{m^2r^2 }{16\pi}\left(\Lambda_g G^{(g)}+\frac{(\beta_1+\omega \beta_2)(G^{(f)}-\omega^2 G^{(g)})(|\Dot{T}|-\omega)}{2\omega^2}\right)~,\\
    & \tilde{T}^{(g)}=-\frac{m^2 }{16\pi}\left[\Lambda_g K^{(g)}-\frac{(\beta_1+\omega \beta_2)}{2\omega^2}\left((K^{(f)}-\omega^2 K^{(g)})(|\Dot{T}|-\omega)+\frac{(H^{(f)}_{AB}-\omega^2 H^{(g)}_{AB})\mathbb{R}^{BA}}{|\Dot{T}|+\omega}\right)\right]~,
\end{align}
with the following two matrices:
\begin{align}
    \mathbb{P}_{AB}&:=\begin{pmatrix}
        \Dot{T}^2(\Sigma_f-\Sigma_g) & \Sigma_f \Dot{T}T^\prime\\
       \vspace{-0.2cm} &\\
        \Sigma_f \Dot{T}T^\prime & \frac{\Dot{T}^2\Sigma_f-\omega^2\Sigma_g}{\Sigma_g^2}
    \end{pmatrix}~,\\
    \mathbb{R}^{AB} & :=\begin{pmatrix}
        \dot{T}^2\dfrac{\Sigma_f-\Sigma_g}{\Sigma_g^2} & -\Sigma_f \dot{T}T^\prime\\
       \vspace{-0.2cm} &\\
        -\Sigma_f \dot{T}T^\prime & \dot{T}^2\Sigma_f-\omega^2\Sigma_g
    \end{pmatrix}~.
\end{align} 
The corresponding source terms for the $f$-sector read as
\begin{align}
  & T^{(f)}_{AB}=-\frac{m^2}{8\pi\alpha^2}\left(\Lambda_f H^{(f)}_{AB}+\frac{(\beta_1+\omega \beta_2)(K^{(f)}-\omega^2 K^{(g)})\mathbb{P}_{AB}}{\omega^3|\Dot{T}|(|\Dot{T}|+\omega)}\right)~,\\
   & T^{(f)}_{A}=-\frac{m^2}{16\pi\alpha^2}\left(2\Lambda_f H^{(f)}_{A}+\frac{(\beta_1+\omega \beta_2)(H^{(f)}_{B}-\omega^2 H^{(g)}_{B})\mathbb{Q}^B_{\,\,\,\,A}}{\omega^3\Sigma_g|\Dot{T}|(|\Dot{T}|+\omega)}\right)~,\\
    & T{(f)}=-\frac{m^2\omega^2r^2 }{16\pi\alpha^2}\left(\Lambda_f G^{(f)}-\frac{(\beta_1+\omega \beta_2)(G^{(f)}-\omega^2 G^{(g)})(|\Dot{T}|-\omega)}{2\omega^5|\Dot{T}|}\right)~,\\
    & \tilde{T}^{(f)}=-\frac{m^2 }{16\pi\alpha^2}\left[\Lambda_f K^{(f)}+\frac{(\beta_1+\omega \beta_2)}{2\omega^5|\Dot{T}|}\left((K^{(f)}-\omega^2 K^{(g)})(|\Dot{T}|-\omega)+\frac{(H^{(f)}_{AB}-\omega^2 H^{(g)}_{AB})\mathbb{R}^{BA}}{|\Dot{T}|+\omega}\right)\right]~.
\end{align}
In the above expressions one can explicitly check that, as commented previously, for the particular case $\beta_1+\omega\beta_2=0$, the metric sectors are decoupled also at the linear level. Finally, we would like to remark that the case $|\dot{T}|=\omega$ analyzed in Ref.~\cite{Babichev:2015zub} is, at first sight, a very particular choice that considerably simplifies the source terms. Even more, as shown in the mentioned reference, for this choice both metrics can be conveniently written in the advanced Eddington-Finkelstein form, simplifying even more the expressions above.

\section{Conclusion} \label{Sec:Conclusion}

We have presented the equations to describe the evolution of linear perturbations of bimetric gravity on a completely general spherically symmetric background spacetime. In order to obtain a covariant setup, valid for any coordinate choice, we have followed the formalism by Gerlach-Sengupta. More precisely, we have performed a 2+2 decomposition of the manifold, so that the background metric is written as a warped product between a two-dimensional metric on a Lorentzian manifold and the metric of the two-sphere. Then we have decomposed all perturbative variables in the natural basis given by tensor spherical harmonics. This removes the dependence on the angles from the different equations and defines two polarity sectors (axial and polar), which evolve independently at the linear level.

In the bimetric theory, there are two sets of equations for linear perturbations, one set for each metric, that couple through effective stress-energy tensors determined by the bimetric interaction potentials. That is, in addition to the contribution of ordinary matter fields, each metric sees the other effectively behaving as a source in the field equations. Hence, the difference with respect to GR, where the matter stress-energy tensor is independently prescribed and matter perturbations are defined independently of the geometry, lies in the fact that here one needs to obtain the explicit expressions for the perturbed effective stress-energy tensors in terms of the perturbations of the two metrics. Such expressions are presented in Sec.~\ref{Sec:First-orderEOM}, and represent one of the main results of this paper.

Owing to the fact that there are twice as many variables as in GR, the dynamical content of the theory is much more intricate and, instead of two, there are seven propagating degrees of freedom. In particular, we have discussed the number of propagating degrees of freedom for each polarity sector and for each multipole $l=0$, $l=1$, and $l\geq 2$. However, the construction of explicit master equations to describe these physical degrees of freedom in the general case is far from trivial. In GR, for a general spherical background, only the Gerlach-Sengupta master equation is known in the axial sector, but there is no such result for the polar one. For the bimetric theory, we have followed the construction by Gerlach-Sengupta for the axial sector and shown that the obstruction to obtain an unconstrained independent equation for the Gerlach-Sengupta master variable lies, on the one hand, in the coupling between the perturbations of the two metrics, and, on the other hand, in the fact that, unlike in GR, there is not enough gauge freedom to remove certain variables.

This formalism is valid for any spherically symmetric background, which, in general, might be dynamical. Even so, as an interesting application, in the last section we have considered the case of a nonbidiagonal static background. More precisely, we have assumed that one of the background metrics ($g_{\mu\nu}$) contains a hypersurface-orthogonal Killing field, and that there is no chart where both metrics are simultaneously diagonal. These assumptions imply that both background geometries are solutions of the Einstein equations, and thus they correspond to the Schwarzschild-(anti)de Sitter geometry with collinear Killing vector fields, while deviations from GR become manifest at the perturbative level. In addition to the two masses and the two cosmological constants, the only freedom at the background level corresponds to the norm of the Killing field of $f_{\mu\nu}$, which is encoded in the function $T=T(t,r)$ that obeys Eq.~\eqref{eq:Tequation}. It is not possible to obtain the general analytic solution for this equation, and thus we have left $T(t,r)$ unspecified in the evolution equations for the perturbations, so as to ensure that our results are valid for any static nonbidiagonal solution and are presented in a form suitable for future studies.

\section*{Acknowledgements}
This work has been supported by the Basque Government Grant 
\mbox{IT1628-22} and by the Grant PID2021-123226NB-I00 (funded by
MCIN/AEI/10.13039/501100011033 and by ``ERDF A way of making Europe'').
ASO acknowledges financial support from the fellowship PIF21/237
of the UPV/EHU. MdC acknowledges support from INFN (iniziative specifiche QUAGRAP and GeoSymQFT). 

\bibliographystyle{bib-style}
\bibliography{references}

\end{document}